\documentclass[aps,prb,twocolumn]{revtex4} 

\usepackage{graphicx}
\usepackage{dcolumn}
\usepackage{bm}

\newcommand{\comment}[1]{}

\begin{document}
\renewcommand{\theequation}{\arabic{section}.\arabic{equation}}

\title{Quantum Statistical Mechanics.
II. Stochastic Schr\"odinger Equation}


\author{Phil Attard}

\date{16 Dec., 2013 (v1). 10 Jan., 2014 (v2). 2 Jun. 2014 (v3).
phil.attard1@gmail.com} 

\begin{abstract}
The stochastic dissipative Schr\"odinger equation
is derived for an open quantum system consisting of a sub-system
able to exchange energy with a thermal reservoir.
The resultant evolution of the wave function
also gives
the evolution of the density matrix,
which is an explicit, stochastic form  of the Lindblad master equation.
A quantum fluctuation-dissipation theorem is also derived.
The time correlation function is discussed.
\end{abstract}


\maketitle

                \section{Introduction}

Quantum mechanics is often portrayed as synonymous
with randomness and unpredictability,
as opposed to the deterministic nature of the classical world.
However this distinction is overly simplistic
since Schr\"odinger's equation is completely deterministic
and it gives the evolution of the wave function of an isolated system
with complete certainty.
Conversely the classical evolution of a sub-system
that can exchange energy with a thermal reservoir
has a random character such that the future can only be predicted
from the present state of the sub-system with statistical probability
rather than deterministic certainty.\cite{TDSM,NETDSM}
In this latter case Hamilton's equations of motion
for the isolated system must be augmented with dissipative and stochastic
terms that arise from the probabilistic treatment of the reservoir.
\cite{NETDSM,AttardIX}
This raises the question of whether one might similarly modify
Schr\"odinger's equation in the quantum case
in order to incorporate the reservoir  environment probabilistically.

The main result obtained in the present paper
is the stochastic dissipative Schr\"odinger equation
for a sub-system able to exchange energy with a thermal reservoir.
This is a direct analogue of the stochastic dissipative
Hamilton's equation that has recently been derived
for classical equilibrium statistical mechanics.
\cite{NETDSM,AttardIX}
The stochastic dissipative Schr\"odinger equation
derived here provides a new way to treat equilibrium quantum systems.
One possible application is as a thermostat
for computer simulation algorithms for quantum statistical mechanics.
In addition, since it gives the evolution of an open quantum system,
it will likely play a role in the development
of a theory for non-equilibrium quantum statistical mechanics.

The origin of the stochastic  Schr\"odinger equation
in the present formulation of quantum statistical mechanics
differs from the literature
(e.g.\ Refs~[\onlinecite{Davies76,Kallianpur80,%
Gisin84,Belavkin89,Kummerer03,Bouten04,Pellegrini08}]).
Also called Belavkin equations,
stochastic  Schr\"odinger equations conventionally
describe the evolution of the quantum state
of a continuously measured system.
In this interpretation,
the stochastic character of the  quantum trajectories
result from the measurement process.

A second, related field of study concerns quantum dissipative systems,
which modify the Schr\"odinger equation
with Langevin-type dissipative and fluctuation terms.\cite{Breuer02,Weiss12}
This approach is related to that  taken here
in that it deals with an open quantum system
with the extra terms accounting for the interactions
with the reservoir beyond the sub-system of direct interest.
In the literature one finds three main approaches:\cite{Weiss12}
modification of the procedure of quantization,
such as using complex variables, or a non-linear Schr\"odinger equation,
\cite{Dekker77,Kostin72,Yasue78,Nelson66}
the  postulation of a heuristic stochastic form
of the Schr\"odinger equation,\cite{Gisin84,Percival98,Wiseman92,Breuer02}
and the projection from the reservoir
onto the sub-system of the density matrix,
\cite{Spohn80,Gardiner91}
or operator.\cite{Gardiner88,Iles-Smith13}
In the projection category one can note the quantum Langevin equation,
\cite{Ford87,Ford88}
which can be derived from an harmonic oscillator bath with linear coupling,
the so-called Caldeira-Leggett model.\cite{Caldeira81}

The stochastic  Schr\"odinger equation
developed here also arises
from the projection of the reservoir onto the sub-system.
Here the emphasis on the statistical origin  of the equation
and the exact symmetry requirements that the equation must obey.
The derivation follows closely that given
in the classical case of the stochastic Hamilton's
equations of motion.\cite{NETDSM,AttardIX}
The present equation allows the computation
of the time evolution of the wave function
of a sub-system of a thermal reservoir,
with the reservoir accounted for in a macroscopic probabilistic sense.
The stochastic and deterministic terms that arise
are shown to be related
by a quantum fluctuation-dissipation theorem.

                \section{Static Properties} \label{Sec:static}
\setcounter{equation}{0}

\subsection{Background} \label{Sec:Stat-Av}

In Paper I,\cite{QSM1}
two results were established that provide the starting point
for the present paper.
For a canonical equilibrium system
(i.e.\ a sub-system that can exchange energy
with a thermal reservoir of temperature $T$),
it was shown that the probability operator had Maxwell-Boltzmann form
\begin{equation} \label{Eq:MB}
\hat \wp = \frac{1}{Z(T)} e^{-\hat{\cal H}/k_\mathrm{B}T} .
\end{equation}
Here $\hat{\cal H}$ is the Hamiltonian or energy operator
of the sub-system and  $k_\mathrm{B}$ is Boltzmann's constant.
This result is the same as the conventional expression
that is postulated for the canonical quantum probability operator.
\cite{Neumann27,Messiah61,Merzbacher70,Bogulbov82}

For future reference the entropy operator is
$\hat S = -\hat{\cal H}/T$.
The normalized energy eigenfunctions $\zeta_n^E$,
$ \hat{\cal H} |\zeta_n^E \rangle = E_n |\zeta_n^E \rangle $,
are also entropy eigenfunctions, $\zeta_n^S=\zeta_n^E$,
$ \hat S |\zeta_n^S \rangle =  S_n |\zeta_n^S \rangle $,
with $S_n \equiv S_{nn}^S =-E_n/T$.
Here the degeneracy is not shown explicitly .

The second result invoked here from Paper I\cite{QSM1}
is that the statistical average of an operator
has the conventional von Neumann form,
\cite{Neumann27,Messiah61,Merzbacher70,Bogulbov82}
\begin{equation} \label{Eq:vN-Tr}
\left< \hat O \right>_\mathrm{stat}
=
\mbox{Tr } \hat \wp \, \hat O .
\end{equation}

These two results followed
from the conservation law for energy,
which  implied that the sub-system wave function
and the reservoir wave function are entangled.\cite{QSM1}
This in turn implied that the principle energy quantum states collapsed.
The degenerate reservoir energy quantum states
were shown to sum to give the Maxwell-Boltzmann form
for the probability operator.
Whilst superposition states with a given energy
can be composed from the degenerate sub-system energy quantum states,
it was shown that these cancel
and that there is no contribution  to the statistical average of an operator
from the non-diagonal terms in the energy representation.
Since the statistical average contains only diagonal terms
in either the energy or the operator representation,
it may be interpreted as resulting from the collapse of the wave function
(c.f. Eqs~(\ref{Eq:Ostat-ne}) and (\ref{Eq:Ostat-eq}) below).

The goal of this paper is to derive a stochastic dissipative
form of the Schr\"odinger equation that represents the evolution
of the wave function of the sub-system in the presence of the reservoir.
Accordingly an expression will be required for the average of an operator
that invokes the resultant wave function and its trajectory.
The focus  is less  on pure quantum states,
as in the von Neumann expression, Eq.~(\ref{Eq:vN-Tr}),
and more on expressing the average as an expectation value
of a wave function state.

Toward this end,
and to illustrate the underlying philosophy of the present approach,
it is now shown that the von Neumann statistical average,
Eq.~(\ref{Eq:vN-Tr}),
can be written as an integral of an expectation value over wave space.
Making an expansion in entropy eigenfunctions, one has
\begin{eqnarray} \label{Eq:O-vN2}
\lefteqn{
\int \mathrm{d} \psi \;
\frac{\langle \psi| \hat \wp \, \hat O| \psi \rangle
}{\langle \psi| \psi \rangle }
} \nonumber \\
& = &
\int \mathrm{d} \underline \psi^S \;
\frac{1}{N(\psi)}
\sum_{lmn} \psi_l^{S*} \psi_n^S \wp^S_{lm} O^S_{mn}
\nonumber \\ & = &
\sum_{mn} \wp^S_{mm} O^S_{mn}
\int \mathrm{d} \underline \psi^S \;
\frac{1}{N(\psi)}  \psi_m^{S*} \psi_n^S
\nonumber \\ & = &
\sum_{n}  \wp^S_{nn} O^S_{nn}
\int \mathrm{d} \underline \psi^S \;
\frac{1}{N(\psi)}
\psi_n^{S*} \psi_n^S
\nonumber \\ & = &
\mbox{const.} \sum_{mn}  \wp^S_{mn} O^S_{nm}
\nonumber \\ & = &
\mbox{Tr } \hat \wp \, \hat O .
\end{eqnarray}
The third equality follows because the integrand for the terms $m\ne n$
is odd, and so they vanish upon integration.
The fourth equality follows because
the value of the integral does not depend upon  $n$,
and because $\wp^S_{mn} = \wp^S_{nn} \delta_{mn}$.
The constant is incorporated into the normalization factor.

Continuing the focus on wave space,
one can explore the possible utility  of a probability density.
For the present canonical equilibrium case,
for which the Maxwell-Boltzmann distribution holds,
there are two plausible definitions namely
\begin{equation}
\wp(\psi) =
\frac{1}{Z(T)}
\frac{ \langle \psi | e^{-\hat{\cal H}/k_\mathrm{B}T} | \psi \rangle
}{  \langle \psi |  \psi \rangle} ,
\end{equation}
or
\begin{equation}
\tilde \wp(\psi) = \frac{1}{Z(T)} e^{-E(\psi)/k_\mathrm{B}T} , \;\;
E(\psi) = \frac{ \langle \psi | \hat{\cal H} | \psi \rangle }{
\langle \psi |  \psi \rangle} .
\end{equation}
These differ if the system is in a superposition of entropy states.
However if the system has collapsed into a pure entropy eigenstate,
then these are equal
$\wp(\zeta_{\alpha g}^\mathrm{S})
= \tilde \wp(\zeta_{\alpha g}^\mathrm{S})
= Z(T)^{-1} e^{-E_\alpha/k_\mathrm{B}T}$.
Neither probability density gives
the statistical average of an operator
as an integral over wave space
\begin{equation} \label{Eq:Ostat-ne}
\langle \hat O \rangle_\mathrm{stat}
\ne
\int \mathrm{d} \psi \; \wp(\psi) \, O(\psi) ,
\end{equation}
where the expectation value of the operator is
$O(\psi) =  \langle \psi | \hat O | \psi \rangle
/ \langle \psi | \psi \rangle$.
The reason that this fails is that this expression always contains
contributions from the superposition states
and never reduces to a sum over pure quantum states
of either the observable operator or of the entropy operator.

However, the collapse
of the wave function into entropy eigenstates yields
\begin{eqnarray} \label{Eq:Ostat-eq}
\left< \hat O \right>_\mathrm{stat}
& = &
\int_\mathrm{coll.} \hspace{-.3cm} \mathrm{d} \psi \;
\wp(\psi)  \, O(\psi)
\nonumber \\ & = &
\sum_n \wp(\zeta_n^S) \, O(\zeta_n^S)
\nonumber \\ & = &
\frac{1}{Z(T)}
\sum_n e^{-E_n/k_\mathrm{B}T}   O_{nn}^S
\nonumber \\ & = &
\mbox{Tr } \hat \wp \,\hat O .
\end{eqnarray}

Whereas $\wp(\psi)$ gives directly the operator $\hat \wp$
and so the von Neumann trace form,
the density $\tilde \wp(\psi)$ does not directly give
a probability operator and the average cannot be written as a trace
over operators (but it can be written as a sum over entropy states
of this probability density).
The `correct' probability density is related to
the `correct' or actual entropy,
$\wp(\psi) = Z(T)^{-1} \exp S_\mathrm{r}(\psi)/k_\mathrm{B}$,
where the entropy is
$S_\mathrm{r}(\psi) = k_\mathrm{B} \ln
\langle \psi | e^{ \hat S_\mathrm{r} /k_\mathrm{B}}  | \psi \rangle
/\langle \psi | \psi \rangle$,
whereas the `incorrect' probability density
is related to the expectation entropy,
$\tilde \wp(\psi) = Z(T)^{-1} \exp S_\mathrm{r}^{<>}(\psi)/k_\mathrm{B}$,
where
$S_\mathrm{r}^{<>} (\psi) =
\langle \psi |  \hat S_\mathrm{r}  | \psi \rangle
/\langle \psi | \psi \rangle$.
\cite{QSM3}

The main focus of the present paper
is on the stochastic, dissipative Schr\"odinger equation
and the trajectory that it generates through wave space.
Accordingly, the relationship between the above expressions
for the statistical average and an expression based upon
the time average over the trajectory is now discussed.

An obvious requirement for an equilibrium system
is that the probability operator be stationary.
With $\hat{\cal U}(t',t)$ the appropriate stochastic, dissipative propagator,
one must have
\begin{equation}
\hat \wp
=
\left<
\hat{\cal U}(t',t)^\dag  \hat \wp \; \hat{\cal U}(t',t)
\right>_\mathrm{stoch},
\end{equation}
where the Maxwell-Boltzmann operator appears on both sides.
This is a formal requirement for the time evolution
of the \emph{equilibrium} probability  operator.

The stochastic, dissipative propagator can also be used to generate
a trajectory in wave space,
\begin{equation}
| \psi(t') \rangle
\equiv
| \psi(t'|\psi,t) \rangle
=
\hat{\cal U}(t',t) \, | \psi \rangle .
\end{equation}
In view of the stationarity of the probability operator
with respect to  the propagator,
one might suppose that this trajectory samples wave space
in proportion to the probability density.
Such a supposition is only partially correct,
as is discussed in the text.

With a single wave function,
one can only make a superposition of states,
not a mixture of states.
So even if the wave function was composed
of entropy modes with the correct entropy weight,
$ | \psi \rangle  =
\sum_n e^{S_n/2k_\mathrm{B}}  | \zeta_n^\mathrm{S}  \rangle$,
the density matrix formed from it would not equal the probability operator,
\begin{equation}
| \psi \rangle \langle \psi |
= \frac{1}{Z(T)} \sum_{nm} e^{[S_m+S_n]/2k_\mathrm{B}}
 | \zeta_m^\mathrm{S}  \rangle \langle \zeta_n^\mathrm{S} |
\ne \hat \wp  .
\end{equation}
(The probability operator is of course diagonal
in the entropy representation.)

One can nevertheless use a density matrix
of the wave function provided that the superposition states average to zero
over the trajectory.
If the wave function evolves without mixing modes,
\begin{equation}
\psi_m^\mathrm{S}(t') = e^{i\phi_m t'} \psi_m^\mathrm{S}(0).
\end{equation}
then the time average of the density matrix is
\begin{eqnarray}
\lefteqn{
\frac{1}{t} \int_0^t \mathrm{d} t'\;
|  \psi(t') \rangle \, \langle \psi(t') |
} \nonumber \\
& = &
\frac{1}{t} \int_0^t \mathrm{d} t'\;
\sum_{mn}
\psi_m^\mathrm{S}(t') \, \psi_n^\mathrm{S}(t')^* \,
 | \zeta_m^\mathrm{S}  \rangle \langle \zeta_n^\mathrm{S} |
\nonumber \\ & = &
\sum_{mn}
\psi_m^\mathrm{S}(0) \, \psi_n^\mathrm{S}(0)^* \,
 | \zeta_m^\mathrm{S}  \rangle \langle \zeta_n^\mathrm{S} |
\nonumber \\ & & \mbox{ } \times
\frac{1}{t} \int_0^t \mathrm{d} t'\;
e^{i(\phi_m - \phi_n)t'}
\nonumber \\ & = &
\sum_{m}
\psi_m^\mathrm{S}(0) \, \psi_m^\mathrm{S}(0)^* \,
 | \zeta_m^\mathrm{S}  \rangle \langle \zeta_m^\mathrm{S} | .
\end{eqnarray}
The off-diagonal phase factors average to zero.
This corresponds to a statistical mixture.
So if the initial coefficients are distributed according
to the exponential of the entropy,
$ \psi_m^\mathrm{S}(0)
\propto  e^{S_m/2k_\mathrm{B}}$,
one will get the correct canonical equilibrium  probability operator
as the time average of the density matrix,
\begin{equation} \label{Eq:wp-MB1}
\hat \wp
=
\frac{1}{t} \int_0^t \mathrm{d} t'\;
\frac{1}{N(\psi(t'))}
| \psi(t') \rangle \, \langle \psi(t')| .
\end{equation}
Here and below, $N(\psi) \equiv \langle \psi | \psi \rangle$ is the norm.

The same result will hold when the wave function mixes entropy modes
as it evolves, since the phase factors of the modes are random.
This case is the most powerful application of the stochastic, dissipative
Schr\"odinger equation discussed in the text
since with it the density matrix converges over the trajectory
to the equilibrium probability operator
from an arbitrary starting wave function.

From this follows the fourth form for the statistical average:
it is a time average of the expectation value
over the stochastic trajectory,
\begin{eqnarray} \label{Eq:<O>-vN-exp}
\left< \hat O \right>_\mathrm{stat}
& = &
\mbox{Tr } \hat \wp \, \hat O
\nonumber \\  & = &
\sum_{mn} \langle \zeta_m | \hat \wp | \zeta_n \rangle \,
\langle \zeta_n | \hat O | \zeta_m \rangle
\nonumber \\  & = &
\frac{1}{t} \int_0^t \mathrm{d} t'\;
\sum_{mn}
\frac{\langle \zeta_m|  \psi(t') \rangle \,
\langle \psi(t')| \zeta_n \rangle \,
O_{nm}
}{N(\psi(t'))}
\nonumber \\  & = &
\frac{1}{t} \int_0^t \mathrm{d} t'\;
\sum_{mn}
\frac{\psi_m(t')  \psi_n(t')^* \, O_{nm}
}{N(\psi(t'))}
\nonumber \\  & = &
\frac{1}{t} \int_0^t \mathrm{d} t'\;
\frac{ \langle \psi(t')| \hat O | \psi(t') \rangle
}{ \langle \psi(t') | \psi(t') \rangle } .
\end{eqnarray}
The final equality has the appearance
that superposition states contribute to the average,
but the first equality makes it clear that these must average to zero.

A direct consequence of the density matrix
averaged over the stochastic trajectory equalling the probability operator,
Eq.~(\ref{Eq:wp-MB1}),
is that different entropy modes are uncorrelated,
\begin{eqnarray} \label{Eq:uncor}
\frac{1}{t} \int_0^t \mathrm{d} t'\;
\frac{\psi_m^S(t')  \psi_n^S(t')^*
}{N(\psi(t'))}
=
\delta_{mn} \wp_{mm}^S .
\end{eqnarray}
This reduces the average to a single sum  in the entropy basis,
as it must.

\subsection{First Entropy}

\subsubsection{Definition}

In the classical case, the exponential of the
entropy is essentially the probability.\cite{TDSM,NETDSM}
A similar definition may be invoked in the quantum case
for the entropy operator,
\begin{equation}
\hat \wp = \frac{1}{Z} e^{\hat S /k_\mathrm{B}}  ,
\end{equation}
where $k_\mathrm{B}$ is Boltzmann's constant.
In view of the form of the Maxwell-Boltzmann probability operator,
Eq.~(\ref{Eq:MB}),
one can identify  the entropy operator
for the canonical equilibrium system as
\begin{equation} \label{Eq:Stot(psi)}
\hat S = \frac{-1}{T} \hat{\cal H} .
\end{equation}
(Logically, the entropy operator is first derived,
and the Maxwell-Boltzmann form for the probability
follows.)\cite{QSM1}

The expectation value of the reservoir entropy
(expectation entropy, for short) is
\begin{equation}
S_\mathrm{r}^{<>}(\psi)
\equiv
\frac{ \langle \psi | \hat S | \psi  \rangle
}{ \langle \psi  | \psi \rangle }
=
\frac{-1}{T}
\frac{ \left< \psi \right| \hat {\cal H}  \left| \psi \right>
}{  \left< \psi | \psi \right> } .
\end{equation}

The actual entropy is
\begin{equation}
S_\mathrm{r}(\psi) =
k_\mathrm{B} \ln
\frac{ \langle \psi | e^{\hat S /k_\mathrm{B}} | \psi  \rangle
}{ \langle \psi  | \psi \rangle } .
\end{equation}
This gives the reservoir entropy for the given wave state of the sub-system.
This is the same as the total entropy,
$S_\mathrm{tot}(\psi) \equiv S_\mathrm{r}(\psi) + S_\mathrm{s}(\psi)
= S_\mathrm{r}(\psi)$,
since the sub-system entropy of a wave state vanishes
(see Appendix B of Ref.~\onlinecite{QSM1}).

As mentioned above,
the `correct' probability density is related to
the `correct' or actual entropy,
$\wp(\psi) = Z(T)^{-1} \exp S(\psi)/k_\mathrm{B}$,
whereas the `incorrect' probability density
is related to the expectation entropy,
$\tilde \wp(\psi) = Z(T)^{-1} \exp S^{<>}(\psi)/k_\mathrm{B}$.

The  projection of the wave function onto the entropy ground state is denoted
$| \overline \psi \rangle = \hat{\cal P}_0 |\psi\rangle
= \sum_g | \zeta_{0 g}^\mathrm{S}  \rangle \,
\langle \zeta_{0 g}^\mathrm{S} |\psi\rangle $.
The  maximum value of the entropy
occurs in the ground state value, and it is denoted
$S_0 \equiv  S_\mathrm{r}(\overline \psi) \equiv \overline S_\mathrm{r}$.
In this sense this can be called the most likely wave state,
although such a designation is a a little misleading
because it is not a unique wave state but rather
the projection of the current wave state onto
the most likely sub-space.

\subsubsection{Derivatives}

The thermodynamic force is
\begin{eqnarray}
\hat S_\mathrm{r}'(\psi) | \psi \rangle
& \equiv &
\frac{\partial S_\mathrm{r}(\psi) }{\partial \langle \psi |}
\nonumber \\ & = &
\frac{k_\mathrm{B}}{N(\psi)} \left[
\frac{ e^{ \hat S_\mathrm{r} /k_\mathrm{B} } }
{ e^{ S_\mathrm{r}(\psi)/k_\mathrm{B} }  }
-  \hat \mathrm{I} \right]  | \psi \rangle .
\end{eqnarray}
This defines the entropy force operator,
$\hat S_\mathrm{r}'(\psi)$,
which, because it depends upon the wave state,
is a non-linear operator.

The second derivative gives
the entropy fluctuation operator,
\begin{eqnarray}
\hat S_\mathrm{r}''(\psi) & \equiv &
\frac{\partial S_\mathrm{r}(\psi)
}{ \partial | \psi \rangle  \partial \langle \psi |}
 \\ & = &
k_\mathrm{B}  \left[
\frac{e^{ \hat S_\mathrm{r} /k_\mathrm{B} }}
{\langle \psi | e^{ \hat S_\mathrm{r} /k_\mathrm{B} } | \psi \rangle }
-
\frac{1}{\langle \psi | \psi \rangle} \hat \mathrm{I}
\right]
\nonumber \\ && \mbox{ }
- k_\mathrm{B}
\frac{e^{ \hat S_\mathrm{r} /k_\mathrm{B} }| \psi \rangle
\langle \psi | e^{ \hat S_\mathrm{r} /k_\mathrm{B} }  }
{\langle \psi | e^{ \hat S_\mathrm{r} /k_\mathrm{B} } | \psi \rangle^2 }
+ k_\mathrm{B}
\frac{| \psi \rangle \langle \psi | }{\langle \psi | \psi \rangle^2 } .
\nonumber
\end{eqnarray}
This is also a non-linear operator.

When the entropy fluctuation operator
is evaluated in the entropy ground state
the final two terms cancel, so that it is equal to the
thermodynamic force operator,
\begin{eqnarray} 
\hat S_\mathrm{r}''(\overline \psi)
 & = &
\hat S_\mathrm{r}'(\overline \psi)
=
\frac{ k_\mathrm{B} }{N(\overline \psi)} \left[
\frac{ e^{ \hat S_\mathrm{r} /k_\mathrm{B} } }
{ e^{ S_0 /k_\mathrm{B} } }
-  \hat \mathrm{I} \right].
\end{eqnarray}
These will also be denoted
$\hat {\overline S}_\mathrm{r}\!'' = \hat {\overline S}_\mathrm{r}\!'$.
The non-linearity means that effectively the entropy gradient
decreases in magnitude with increasing projection of the wave function
onto the ground state.
By design, the thermodynamic force vanishes
in the most likely wave state,
$\hat {\overline S}_\mathrm{r}\!'' | \overline \psi \rangle
= \hat {\overline S}_\mathrm{r}\!' | \overline \psi \rangle
= | 0 \rangle$.
Further, the fluctuation operator is negative semi-definite,
$\langle \psi | \hat {\overline S}_\mathrm{r}\!'' | \psi \rangle
\le 0$.

\subsubsection{Fluctuation Form}  
\label{Sec:S1-flucn}

As mentioned above, the total entropy is given by
\begin{equation}
e^{S(\psi|T)/k_\mathrm{B}}
=
\frac{  \left< \psi \right| e^{\hat S_\mathrm{r}/k_\mathrm{B}}
 \left| \psi \right>
}{ \left< \psi | \psi \right> } .
\end{equation}
In the canonical equilibrium case
the entropy operator is $\hat S_\mathrm{r} = - \hat{\cal H}/T$.

One may define the fluctuation of a wave function
as its departure from the most likely value,
$\Delta \psi \equiv  \psi - \overline \psi$.
In terms of this, one carries out a second order expansion
about the most likely wave function
and writes the first entropy in fluctuation approximation as
\begin{equation} \label{Eq:S1-flucn}
S_\mathrm{r}(\psi) =
\overline S_\mathrm{r}
+
\left< \Delta \psi \right| \hat {\overline S}_\mathrm{r}\!''
 \left| \Delta \psi \right>
 + {\cal O}(\Delta \psi^3),
\end{equation}
where the value in the most likely state is
$ \overline S_\mathrm{r} \equiv S_\mathrm{r}(\overline \psi)
= S_0 = - E_0/T$.
As mentioned, because the most likely state maximizes the first entropy,
the entropy fluctuation operator is negative semi-definite.

Since $ \hat {\overline S}_\mathrm{r}\!''
= \hat {\overline S}_\mathrm{r}\!'$,
differentiating the fluctuation form gives
the thermodynamic force as
${\partial S_\mathrm{r}(\psi) }/{\partial \langle \psi |}
=
\hat {\overline S}_\mathrm{r}\!'' | \Delta \psi \rangle
+ {\cal O}(\Delta \psi^2)$.
One could replace in this $\hat {\overline S}_\mathrm{r}\!''
\Rightarrow \hat {S}_\mathrm{r}\!'(\psi)$,
or  $  | \Delta \psi \rangle \Rightarrow  |  \psi \rangle$,
or both,
without changing the leading order in the fluctuation expansion.
However, in this case the fluctuation operator is no longer
negative semi-definite, which is a disadvantage.

\subsection{Helmholtz Free Energy} \label{Sec:Helmholtz}

The total unconstrained entropy
is the logarithm of the  partition function,
and it may be rewritten as
\begin{eqnarray}
S_\mathrm{tot}(T) & = &
k_\mathrm{B} \ln Z(T)
\nonumber \\ & = &
k_\mathrm{B} \mbox{Tr} \left\{ \hat \wp(T)  \ln Z(T) \right\}
\nonumber \\ & = &
k_\mathrm{B} \mbox{Tr} \left\{ \hat \wp(T) \left[
- \ln \hat \wp(T)
+ \ln  e^{- \hat{\cal H} /k_\mathrm{B}T} \right]  \right\}
\nonumber \\ & = &
\frac{-1}{T} \mbox{Tr} \left\{ \hat \wp(T) \, \hat{\cal H}  \right\}
\nonumber \\ &  & \mbox{ }
- k_\mathrm{B} \mbox{Tr} \left\{ \hat \wp(T) \ln \hat \wp(T)   \right\} .
\end{eqnarray}
This is written in the form of a sum over quantum states.
It can be equivalently written as
an integral over the Hilbert space,
\begin{eqnarray}
S_\mathrm{tot}(T)
& = &
k_\mathrm{B} \ln Z(T)
 \\ \nonumber & = &
k_\mathrm{B}
\int  \mathrm{d} \psi  \,
\frac{ \left< \psi \right| \hat \wp( T)  \ln Z(T)  \left| \psi \right>
}{\left< \psi | \psi \right>}
\nonumber \\ & = &
k_\mathrm{B}
\int  \mathrm{d} \psi  \,
\left[
\frac{ - \left<\psi \right|   \hat \wp(T)\ln \hat \wp(T)
\left|  \psi \right>
}{\left< \psi | \psi \right>}
\right. \nonumber \\ && \mbox{ } \left.
+
\frac{ \left< \psi \right|
\hat \wp(T) \ln e^{- \hat{\cal H} /k_\mathrm{B}T}
\left| \psi \right>
}{\left< \psi | \psi \right>}
\right]
\nonumber \\ & = &
\frac{- \left< \hat{\cal H}  \right>_{T} }{T}
-k_\mathrm{B}
\int \mathrm{d} \psi  \,
\frac{ \left< \psi  \right| \hat \wp(T) \ln \hat \wp(T)
\left| \psi \right>
}{\left< \psi | \psi \right>}  .
\nonumber 
\end{eqnarray}
The equivalence of these expressions
follows from the collapse of the wave function,
Eq.~(\ref{Eq:O-vN2}):
the trace can be formulated equivalently as a sum over quantum states
or as an integral over Hilbert space.

The first term is evidently the unconstrained reservoir entropy
(c.f.\ Eq.~(\ref{Eq:Stot(psi)}), averaged over the sub-system microstates),
\begin{equation}
S_\mathrm{r}(T)
=
\frac{- 1 }{T} \left< \hat{\cal H}    \right>_{T} .
\end{equation}
This is the average of wave space
of the expectation entropy.
In so far as the wave function has collapsed
into entropy eigenstates,
the latter is the same as the actual entropy.

Because the probabilities are sharply peaked,
canonical equilibrium averages
are equal to micro-canonical or isolated equilibrium averages.
Accordingly, the remaining term is the unconstrained sub-system entropy,
\begin{eqnarray} \label{Eq:S_s-unconstr}
S_\mathrm{s}(T)
& = &
- k_\mathrm{B} \left< \ln \hat \wp(T) \right>_{T}
\nonumber \\ & = &
- k_\mathrm{B} \mbox{Tr} \left\{ \hat \wp(T) \ln \hat \wp(T)   \right\}
\nonumber \\ & = &
-k_\mathrm{B}
\int \mathrm{d} \psi  \,
\frac{ \left< \psi \right|
\hat \wp(T) \ln \hat \wp(T)
\left| \psi \right>
}{\left< \psi | \psi \right> } .
\end{eqnarray}

This formula for the sub-system entropy is of the form,
$S=-k_\mathrm{B} \sum_\alpha \wp_\alpha \ln \wp_\alpha$,
which is commonly called the information entropy.
In the classical case it is variously attributed to
Boltzmann, Gibbs, and Shannon, and it was popularized by Jaynes.
In quantum mechanics it is called the von Neumann entropy,
and it is written in terms of the density matrix,
$S=-k_\mathrm{B} \mathrm{Tr}  \rho \ln \rho $.
\cite{Messiah61,Merzbacher70,Bogulbov82} 
It is commonly called `the' entropy of the system,
which implies that it is the entropy of the total system,
when in fact it is only the sub-system part of the total entropy.
Because most workers mistake this for the total entropy,
it is commonly maximized to obtain equilibrium properties
including the equilibrium probability distribution
(e.g.\ Jaynes' maxent approach to statistical mechanics).
The consequences and problems
with this common misinterpretation are detailed elsewhere.\cite{Attard12}

One has to distinguish between the statistical mechanical
and the thermodynamic definitions of the free energy.\cite{TDSM}
In  statistical mechanics,
the free energy corresponds to the unconstrained total entropy,
whereas in thermodynamics the free energy corresponds to the
maximal constrained total entropy.
In the thermodynamic limit of an infinitely large sub-system
and relatively negligible fluctuations, the two are equal.
In statistical mechanics,
in general the free energy is minus the temperature
times the total entropy.
For the present  case
of a sub-system exchanging energy with a thermal reservoir
the statistical mechanical Helmholtz free energy is
\begin{eqnarray}
F_\mathrm{SM}(T) & = &
- T S_\mathrm{tot}(T)
\nonumber \\ & = &
-k_\mathrm{B} T \ln Z(T)
\nonumber \\ & = &
\left<  \hat{\cal H}     \right>_{T}
- T S_\mathrm{s}(T).
\end{eqnarray}
This is the obvious analogue of the classical form
for the Helmholtz free energy
that one sees in standard thermodynamic texts.
The main point to note is that it is the sub-system entropy that appears
explicitly.
All text books apart from the author's\cite{TDSM,NETDSM}
call this term `the' entropy,
and imply that it is the entropy of the total system,
which it clearly isn't.
The sub-system entropy that appears here
is that of an unconstrained isolated system,
Eq.~(\ref{Eq:S_s-unconstr}).
The second point to note that it is the average of the normalized energy
that appears;
this is extensive with the sub-system size,
as is
the sub-system entropy.

The thermodynamic  free energy
is minus the temperature times
the maximum value of the constrained total entropy.
By definition the most likely state
maximizes the constrained total entropy.
The most likely
values of the energy and norm may be denoted by an overbar.
In so far as the fluctuations are Gaussian,
these are equal to the average values.

For the present canonical case,
the  thermodynamic  Helmholtz free energy is
\begin{eqnarray}
F_\mathrm{TD}(T) & = &
- T S_\mathrm{tot}(\overline {E} |T)
\nonumber \\ & = &
\overline {E}(T)
- T S_\mathrm{s}(\overline { E}(T)) .
\end{eqnarray}
The sub-system entropy that appears here
is the constrained one
given by the isolated system
with specified values of energy. 
Note the difference between the most likely sub-system entropy,
that appears here $S_\mathrm{s}(\overline { E}(T))$,
(this is approximately the unconstrained sub-system entropy,
and is the sub-system entropy in the most likely sub-system energy
macrostate),
and the reservoir entropy in the most likely wave state
that was used above,
$\overline S_\mathrm{r} \equiv S_\mathrm{r}(\overline \psi)
= S_0 = -E_0/T$.
Note also that
$S_\mathrm{s}(\overline { E}(T)) \ne -\overline { E}(T))/T$,
and that $\overline { E} \ne E(\overline \psi) \equiv E_0$.

There is an obvious identity between the functional forms
for the thermodynamic and statistical mechanical
Helmholtz free energies.
The free energy is negative,
assuming that the constant has been chosen to make the total entropy
positive.
The thermodynamic free energy is strictly greater
(i.e.\ strictly smaller in magnitude)
than the  statistical mechanical free energy.
In the thermodynamic limit of an infinitely large sub-system
and relatively negligible fluctuations, the two are equal
to a very good approximation.

%
%


%
\section{Transition Probability
and the Stochastic Dissipative Schr\"odinger Equation}
\label{Sec:Stoch-EoM}
\setcounter{equation}{0}
%

The aim of this section is to derive the transition probability operator
for a canonical equilibrium system,
and from this to derive the stochastic dissipative Schr\"odinger equation.
The strategy will follow closely the analogous derivation
for the classical case.\cite{NETDSM}

This section is primarily concerned with the wave state transition,
$\left| \psi_1 \right> \stackrel{\tau}{\rightarrow} \left| \psi_2 \right> $.
The time interval $\tau$ can be positive or negative
and is initially of arbitrary duration;
rather quickly a small $\tau$  expansion will be performed.
One can consider the initial and final sub-system
wave functions of the transition
as belonging to distinct Hilbert spaces and form the direct product
${\mathrm H}_1 \otimes {\mathrm H}_2$,
with the total wave function written as
$ \left| \psi_1 \right> \left| \psi_2 \right>
\equiv \left| \psi_1 , \psi_2 \right>$.

The analysis is based upon the second entropy operator,
which may also be called the transition entropy operator,
or the two-time entropy operator.
For the transition in time $\tau$,
the unconditional  transition probability operator
is related to the second entropy operator by
\begin{equation} \label{Eq:hat-wp2}
\hat \wp^{(2)}(\tau)
=
\frac{e^{ \hat S^{(2)}(\tau) /k_\mathrm{B} }
}{Z^{(2)}(\tau,T) } .
\end{equation}
A detailed analysis of the second entropy
in  fluctuation approximation follows shortly.
First, the time symmetries for an equilibrium system will be given explicitly.

\subsection{Time Symmetry of the Adiabatic Propagator} \label{Sec:Time-Sym}

There are two distinct time symmetries that result
from Schr\"odinger's equation for a time-independent isolated quantum system,
namely time reversibility and microscopic reversibility.
In order to derive these one begins by noting
that complex conjugation is the quantum operation
that corresponds to the classical notion of velocity reversal.
To see this,
integrating Schr\"odinger's equation,
\begin{equation}
i \hbar | \dot \psi\,\!^0 \rangle
= \hat{\cal H}  |  \psi \rangle ,
\end{equation}
gives the adiabatic time propagator as
\begin{equation}
\hat U^0(\tau)
\equiv
e^{ \tau \hat{\cal H} /i \hbar } .
\end{equation}
By inspection one sees that there are two symmetries:
The first symmetry is time reversibility,
\begin{equation}
\hat U^0(-\tau)  = \hat U^0(\tau)^{-1} ,
\end{equation}
and,
since the Hamiltonian is  Hermitian,
$\hat{\cal H}^\dag  = \hat{\cal H}$,
and real $\hat{\cal H}^*  = \hat{\cal H}$,
the second symmetry is microscopic reversibility,
\begin{equation}
\hat U^0(\tau)^\dag = \hat U^0(\tau)^* = \hat U^0(\tau)^{-1}.
\end{equation}
Time reversibility says that the trajectory through any point is unique.
Microscopic reversibility says that reversing the velocities
reverses the transition,
which is to say that $\psi_1 \stackrel{\tau}{\rightarrow} \psi_2$
has reverse transition $\psi_2^* \stackrel{\tau}{\rightarrow} \psi_1^*$.
In general, in consequence of microscopic reversibility
the  propagator is unitary.
Note that time is \emph{not} reversed in microscopic reversibility.

It should be emphasized that microscopic reversibility
implies the dual symmetry
$\hat U^0(\tau)^\dag =  \hat U^0(\tau)^{-1}$
and $\hat U^0(\tau)^* = \hat U^0(\tau)^{-1}$.
By the definition that the Hermitian conjugate
is the transpose of the complex conjugate,
$\dag \equiv *\mathrm{T}$,
these together imply the transpose symmetry of the propagator,
\begin{equation}
\hat U^0(\tau)^\mathrm{T}= \hat U^0(\tau).
\end{equation}

The reason why the double symmetry
is implied by microscopic reversibility for a general equilibrium system
is now discussed.
Consider the evolution of the wave function,
\begin{equation}
|  \psi_2 \rangle = \hat U^0(\tau)  |  \psi_1 \rangle .
\end{equation}
The complex conjugate of a wave state is $\psi^*$,
or $ |  \psi \rangle^* = |  \psi^* \rangle$.
Hence the evolution equation has complex conjugate
\begin{eqnarray}
|  \psi_2^* \rangle
& = & \hat U^0(\tau)^*  |  \psi_1^* \rangle
= \hat U^0(\tau)^{-1}  |  \psi_1^* \rangle ,
\nonumber \\
\mbox{ or }\;
|  \psi_1^* \rangle & = & \hat U^0(\tau)  |  \psi_2^* \rangle
\end{eqnarray}
This says that if $\psi_2 $ is the end result of the transition
from $\psi_1$ in time $\tau$,
then  $\psi_1^* $ is the end result of the transition
from $\psi_2^*$, also in time $\tau$.
This has the obvious meaning of reversing the transition,
and that $\psi^*$ is the same as the state $\psi$
with all of the velocities reversed.

The complementary approach
is to take the Hermitian conjugate of the same transition,
\begin{eqnarray}
\langle \psi_2 |  &=&
\langle \psi_1 | \hat U^0(\tau)^\dag
= \langle \psi_1 | \hat U^0(\tau)^{-1} ,
\nonumber \\
\mbox{ or }\;
\langle \psi_1 |   &=& \langle \psi_2 | \hat U^0(\tau)   .
\end{eqnarray}
One can also see this as reversing the transition,
with the interpretation that
$\langle \psi |$
is the dual of the state $ |  \psi \rangle $
with all the velocities reversed.

For an open equilibrium quantum system,
microscopic reversibility must still hold,
because the stochastic, dissipative transitions
of the sub-system are derived from adiabatic transitions of the total system.

\subsection{Second Entropy Fluctuation Operator}

\subsubsection{Fluctuation Form} \label{Sec:S2-flucn}

Recall that the fluctuation of the sub-system wave function is
$\Delta \psi \equiv \psi - \overline \psi$.
Assume that the second entropy for the transition
has the quadratic fluctuation form
\begin{eqnarray} \label{Eq:S2-flucn}
\lefteqn{
S^{(2)}( \psi_2, \psi_1;\tau)
}  \\ \nonumber
& = &
\left< \Delta \psi_2 , \Delta \psi_1 \right|
\hat {\cal A}^{(2)}(\tau)
\left|\Delta \psi_1 , \Delta \psi_2 \right>
+ \overline S^{(1)}
\\ \nonumber & = &
\left< \Delta \psi_2 \right|  \hat a(\tau) \left| \Delta \psi_2 \right>
+
\left< \Delta \psi_1 \right|  \hat c(\tau) \left| \Delta \psi_1 \right>
 \\ \nonumber &&  \mbox{ }
+
\left< \Delta \psi_2 \right|  \hat b(\tau) \left| \Delta \psi_1 \right>
+
\left< \Delta \psi_1 \right|  \hat b(\tau)^\dag \left| \Delta \psi_2 \right>
+ \overline S^{(1)} .
\end{eqnarray}
Note that the order of the arguments of the second entropy
defines the time interval as the time of the first argument
minus the time of the second argument,
$S^{(2)}( \psi_2, \psi_1;\tau) \Rightarrow \tau \equiv t_2 - t_1$.
The various operators are also a function of temperature,
which is generally not shown.
The final (immaterial) constant makes the maximum value of the second entropy
equal to that of the first entropy $\overline S^{(1)}$
when the two termini are the most likely state,
$\psi_1 = \psi_2 = \overline \psi$,
which is related to the reduction condition
discussed further below.\cite{NETDSM,AttardII}
The negative definite operator $\hat {\cal A}^{(2)}(\tau)$
(equivalently its component operators
$\hat a(\tau) $, $\hat b(\tau) $, and $\hat c(\tau) $)
is the second entropy fluctuation operator,
and, after its properties have been established,
it will be used to give the second entropy operator itself,
and hence the transition probability operator.

Three symmetries must hold:
statistical symmetry,
\begin{equation} \label{Eq:S2-stat-sym}
S^{(2)}( \psi_2, \psi_1;\tau)
 =
S^{(2)}( \psi_1, \psi_2;-\tau) ,
\end{equation}
reality,
\begin{equation}
S^{(2)}( \psi_2, \psi_1;\tau) =
S^{(2)}( \psi_2, \psi_1;\tau)^*,
\end{equation}
and microscopic reversibility,
\begin{equation} \label{Eq:mu-rev}
S^{(2)}( \psi_2, \psi_1;\tau) =
S^{(2)}( \psi_1^*, \psi_2^*;\tau) .
\end{equation}
(Note that this means the conjugate of the fluctuation,
$\Delta \psi^* \equiv \psi^* - \overline \psi\,\!^*
= [ \hat{\mathrm I} - \hat{\cal P}_0] | \psi^* \rangle$.)

Statistical symmetry simply reflects the ordering of the arguments
of the second entropy discussed above.
It can also be called time homogeneity symmetry,
since it follows by first shifting the time origin by $t$,
$t_2=t+\tau$ and $t_1 = t$,
and then setting $t \Rightarrow -\tau$.
The second entropy is real for the same reasons
that the first entropy is real:
because it is a physical observable
(more precisely,
a linear combination of physical observables with real coefficients),
and also because its exponential gives the transition probability density,
which is taken to be real.
Microscopic reversibility is non-trivial and relies upon two facts:
First, Schr\"odinger's equation for the total isolated system
(sub-system plus reservoir) obeys microscopic reversibility.
And second, for an equilibrium system,
a wave state of the reservoir $\psi_\mathrm{r}$
and its conjugate $\psi_\mathrm{r}^*$ are equally probable,
since conjugation represents velocity reversal.

Statistical symmetry implies that
\begin{equation}
\hat a(-\tau) = \hat c(\tau)
,\mbox{ and }
\hat b(-\tau)
= \hat b(\tau)^\dag .
\end{equation}
Reality implies that
\begin{equation}
\hat a(\tau)^\dag = \hat a(\tau)
,\mbox{ and }
\hat c(\tau)^\dag = \hat c(\tau) .
\end{equation}
The form of the two cross terms in the second entropy guarantees reality
for the contribution from these two terms.
Microscopic reversibility implies that
\begin{equation}
\hat a(\tau)^* = \hat c(\tau)
,\mbox{ and }
\hat b(\tau)^*   = \hat b(\tau)^\dag .
\end{equation}
Hence $\hat a(\tau) = \hat a(\tau)^\dag = \hat a(-\tau)^* $,
and $\hat b(\tau) = \hat b(\tau)^{*\dag} = \hat b(-\tau)^* $.

\subsubsection{Most Likely Terminus}

The derivative of the second entropy with respect to
$\left< \psi_2 \right| $ is
\begin{equation}
\frac{ \partial S^{(2)}( \psi_2, \psi_1;\tau)
}{\partial \left< \psi_2 \right| }
 =
\hat a(\tau) \left| \Delta \psi_2 \right>
+
\hat b(\tau)  \left| \Delta \psi_1 \right> .
\end{equation}
The most likely terminus of the transition,
$\overline \psi_2 \equiv \overline \psi(\tau|\psi_1)$,
follows by setting this derivative to zero,
\begin{equation}
 \left| \Delta \overline \psi_2 \right>
=
- \hat a(\tau)^{-1} \hat b(\tau)  \left| \Delta \psi_1 \right> .
\end{equation}

\subsubsection{Reduction Condition}

The second entropy may be re-written in terms of the departure
from the most likely terminus,
$\Delta \psi_2 - \Delta \overline \psi_2
= \psi_2 - \overline \psi_2$,
\begin{eqnarray}
\lefteqn{
S^{(2)}( \psi_2, \psi_1;\tau)
= }
\\ \nonumber & &
\left< \psi_2 - \overline \psi_2 \right|
\hat a(\tau)
\left| \psi_2 - \overline \psi_2 \right>
\\  \nonumber && \mbox{ }
+
\left< \Delta \psi_1 \right|
\left\{ \hat c(\tau)
- \hat b(\tau)^\dag \hat a(\tau)^{-1}  \hat b(\tau) \right\}
\left| \Delta \psi_1\right>
+ \overline S^{(1)}.
\end{eqnarray}

The reduction condition is that in the most likely state,
the second entropy reduces to the first entropy,\cite{NETDSM,AttardII}
\begin{eqnarray}
S^{(2)}(\overline \psi_2, \psi_1 ;\tau)
& = &
S^{(1)}( \psi_1 |T)
\nonumber \\  & = &
\left< \Delta \psi_1 \right| \hat S'' \left| \Delta \psi_1 \right>
+ \overline S^{(1)} .
\end{eqnarray}
The reduction condition is equivalent to insisting
that the transition probability
must reduce to the probability of the initial state
upon summing over all possible final states.\cite{NETDSM,AttardII}
This is the reason that it is the second derivative of the actual entropy
and not the expectation entropy that is required for the fluctuation operator.
The reduction condition therefore yields the requirement
\begin{equation}
\hat c(\tau)
- \hat b(\tau)^\dag \hat a(\tau)^{-1}  \hat b(\tau)
= \hat {\overline S}_\mathrm{r}\!'' .
\end{equation}
This result must hold for each value of the time step $\tau$.

\subsubsection{Small Time Expansion}

Since $ \overline \psi_2 \rightarrow \psi_1$ as $\tau \rightarrow 0$,
the second entropy must contain essentially a $\delta$-function singularity.
Hence the small-$\tau$ expansions must be of the form
\begin{equation}
\hat a(\tau) =
\frac{1}{|\tau|} \hat a_{-1} + \frac{1}{\tau} \hat a_{-1}'
+
\hat a_{0} + \widehat \tau \hat a_{0}' + {\cal O}(\tau) ,
\end{equation}
\begin{equation}
\hat b(\tau) =
\frac{1}{|\tau|} \hat b_{-1} + \frac{1}{\tau} \hat b_{-1}'
+
\hat b_{0} + \widehat \tau \hat b_{0}' + {\cal O}(\tau) ,
\end{equation}
and
\begin{equation}
\hat c(\tau) =
\frac{1}{|\tau|} \hat c_{-1} + \frac{1}{\tau} \hat c_{-1}'
+
\hat c_{0} + \widehat \tau \hat c_{0}' + {\cal O}(\tau) ,
\end{equation}
with $\widehat \tau \equiv \mbox{sign }\tau = \tau/|\tau|$.

The  reason why the non-analytic terms appear
(i.e.\ those containing $|\tau|$ and $\widehat \tau$)
is that these are necessary to yield the irreversible behavior
that is characteristic of all thermodynamic evolution.
One concludes that this is not a Taylor expansion
for an infinitesimal time step,
since this would only ever yield analytic terms,
but rather an expansion for small but finite time steps
that is a re-summation of an infinite order  Taylor expansion.
The validity of beginning the expansion
with terms ${\cal O} (\tau^{-1})$ can be judged by the consequences;
amongst other things it yields a physically plausible
stochastic Schr\"odinger equation with a conventional velocity
for the wave function.

From the symmetries given above,
$\hat a(\tau) = \hat a(\tau)^\dag = \hat a(-\tau)^* $,
and $ \hat a(-\tau) = \hat c(\tau)$,
one can see that
the unprimed $\hat a$ are real and self-adjoint
and equal the unprimed $\hat c$,
and the primed $\hat a$ are imaginary  and self-adjoint
and equal the negative of the primed $\hat c$.
Also,
since $\hat b(\tau) = \hat b(\tau)^{*\dag} = \hat b(-\tau)^* $,
the unprimed $\hat b$ are real and self-adjoint,
and the primed $\hat b$ are imaginary and anti-self-adjoint.

Since $ \overline \psi_2 \rightarrow \psi_1$ as $\tau \rightarrow 0$,
to leading order $ \hat a(\tau) = - \hat b(\tau)$,
which implies that
\begin{equation}
\hat a_{-1} = - \hat b_{-1}  \equiv -\hat \lambda^{-1}
, \mbox{ and }
\hat a_{-1}' = -\hat b_{-1}' = \hat 0 .
\end{equation}
From the symmetry relations,
$\hat \lambda$ is a real Hermitian operator that is positive definite
(because the second entropy must be negative definite).
The primed coefficients individually  vanish because
$\hat a_{-1}'$ is self-adjoint
and $\hat b_{-1}'$ is anti-self-adjoint.
With these, the small time expansions  read
\begin{equation}
\hat a(\tau) =
\frac{ - 1}{|\tau|} \hat \lambda^{-1}
+ \hat a_{0} + \widehat \tau \hat a_{0}' + {\cal O}(\tau) ,
\end{equation}
\begin{equation}
\hat b(\tau) =
\frac{1}{|\tau|} \hat \lambda^{-1}
+ \hat b_{0} + \widehat \tau \hat b_{0}' + {\cal O}(\tau),
\end{equation}
and
\begin{equation}
\hat c(\tau) =
\frac{-1}{|\tau|} \hat \lambda^{-1}
+ \hat a_{0} - \widehat \tau \hat a_{0}' + {\cal O}(\tau) .
\end{equation}

Inserting these expansions in the reduction condition,
to zeroth order in the time step one must have
\begin{eqnarray}
\hat {\overline S}_\mathrm{r}\!''
& = &
\hat c(\tau)
- \hat b(\tau)^\dag \hat a(\tau)^{-1}  \hat b(\tau)
\nonumber \\ & = &
\frac{-1}{|\tau|} \hat \lambda^{-1}
+ \hat a_{0} - \widehat \tau \hat a_{0}'
- \left[
\frac{1}{|\tau|} \hat \lambda^{-1}
+ \hat b_{0} - \widehat \tau \hat b_{0}'
\right]
\nonumber \\ &  & \mbox{ } \times
\left[
\frac{ - 1}{|\tau|} \hat \lambda^{-1}
+ \hat a_{0} + \widehat \tau \hat a_{0}'
\right]^{-1}
\left[
\frac{1}{|\tau|} \hat \lambda^{-1}
+ \hat b_{0} + \widehat \tau \hat b_{0}'
\right]
\nonumber \\ & = &
\frac{-1}{|\tau|} \hat \lambda^{-1}
+ \hat a_{0} - \widehat \tau \hat a_{0}'
+ \left[
\frac{1}{|\tau|} \hat \lambda^{-1}
+ \hat b_{0} - \widehat \tau \hat b_{0}'
\right]
\nonumber \\ &  & \mbox{ } \times
\left[ \hat{\mathrm I}
+ |\tau| \hat \lambda \hat a_{0}
+  \tau \hat \lambda \hat a_{0}'
\right]
\left[
\hat{\mathrm I} +
|\tau| \hat \lambda \hat b_{0} +   \tau  \hat \lambda\hat b_{0}'
\right]
\nonumber \\ & = &
\hat a_{0} - \widehat \tau \hat a_{0}'
+ \hat b_{0} - \widehat \tau \hat b_{0}'
+  \hat a_{0} +  \widehat \tau \hat a_{0}'
+  \hat b_{0} +  \widehat \tau \hat b_{0}'
\nonumber \\ & = &
2 \left[ \hat a_{0} + \hat b_{0} \right]
+ {\cal O}(\tau).
\end{eqnarray}
In short,
\begin{equation}
\hat a_{0} + \hat b_{0}
= \frac{1}{2 } \hat {\overline S}_\mathrm{r}\!''
= \frac{1}{2 } \hat {\overline S}_\mathrm{r}\!' .
\end{equation}
This is formally the same as the result for classical fluctuations
in macrostates or microstates given in Ref.~\onlinecite{NETDSM}.

\subsubsection{Dissipative Schr\"odinger Equation}

Using this, the expansion
for the most likely terminal wave function departure becomes
\begin{eqnarray}
 \left| \Delta \overline \psi_2 \right>
& = &
- \hat a(\tau)^{-1} \hat b(\tau)  \left| \Delta \psi_1 \right>
 \\ \nonumber & = &
\left[ \hat{\mathrm I}
+ |\tau| \hat \lambda \hat a_{0}
+  \tau \hat \lambda \hat a_{0}'
\right]
\nonumber \\ &  & \mbox{ } \times
\left[
\hat{\mathrm I} +
|\tau| \hat \lambda \hat b_{0} +  \tau  \hat \lambda \hat b_{0}'
\right] \left| \Delta \psi_1 \right>
\nonumber \\ & = &
\left| \Delta \psi_1 \right>
+
\tau  \hat \lambda \left[ \hat a_{0}' + \hat b_{0}' \right]
\left| \Delta \psi_1 \right>
\nonumber \\ &  & \mbox{ }
+ \frac{ |\tau| }{2 } \hat \lambda
\hat {\overline S}_\mathrm{r}\!'
\left| \Delta \psi_1 \right>
+ {\cal O}(\tau^2) .
\nonumber
\end{eqnarray}
Here the thermodynamic force operator has been used,
$\hat {\overline S}_\mathrm{r}\!'
= \hat {\overline S}_\mathrm{r}\!''$.

It is possible that
variational procedures based upon the approximation
$ \hat {\overline S}_\mathrm{r}\!' \approx
\hat {S}_\mathrm{r}\!'(\psi) + {\cal O}(\Delta \psi)$
might prove useful,
although this does destroy the negative semi-definiteness
of the entropy fluctuation operator (see below).
In fact it is probably not too hard to estimate
$\hat {\overline S}_\mathrm{r}\!'$
since the required $S_0 \equiv  S_{\mathrm{r}}(\overline \psi)$
can be taken to be the maximum value of $ S_{\mathrm{r}}(\psi)$
calculated thus far on the trajectory,
and one can probably make the replacement
$N(\overline \psi) \Rightarrow N(\psi)$,
or else incorporate this constant into the dissipative operator.

The adiabatic evolution must be contained in the reversible term,
which is the one that is proportional to $\tau$,
$| \dot \psi_1^0 \rangle
= (1/i\hbar) \hat {\cal H}  | \psi_1 \rangle$.
There may be reversible reservoir contributions,
but since the reservoir is nothing but a perturbation
to the adiabatic evolution,
they can be neglected compared to the reversible adiabatic term.
One cannot neglect the irreversible reservoir  term
because there is not adiabatic irreversible term with which to compare it.
Since it is the only irreversible term,
and since irreversibility is an essential ingredient in the evolution
that follows directly from the second law of thermodynamics,
one must retain  the irreversible reservoir  term.

Hence equating the reversible term here with the adiabatic evolution
of the wave function, one must have
\begin{equation} \label{Eq:rev-adiabatic+res}
\hat \lambda \left[ \hat a_{0}' + \hat b_{0}' \right]
=
\frac{1}{i\hbar} \hat {\cal H}   .
\end{equation}

With $| \Delta \psi \rangle \equiv
|\psi \rangle -  \hat{\cal P}_0 |\psi \rangle $,
$\overline \psi_2 \equiv \overline \psi(t_2|\psi_1,t_1)$,
and using this result for the reversible evolution,
the above expression for the evolution of the fluctuation
may be rearranged to give the evolution of the wave function itself
with the ground state projection terms explicit,
\begin{eqnarray}
 \left| \overline \psi_2 \right>
& = &
| \psi_1 \rangle
+ \frac{\tau}{i\hbar} \hat {\cal H} | \psi_1 \rangle
+ \frac{ |\tau| }{2 } \hat \lambda
 \hat {\overline S}_\mathrm{r}\!' \, | \psi_1 \rangle
\nonumber \\ &  & \mbox{ }
+  \hat{\cal P}_0 | \overline \psi_2  \rangle
-  \hat{\cal P}_0 | \psi_1  \rangle
- \frac{\tau}{i\hbar}  \hat {\cal H} \hat{\cal P}_0 | \psi_1 \rangle
\nonumber \\ &  & \mbox{ }
- \frac{ |\tau| }{2 } \hat \lambda
 \hat {\overline S}_\mathrm{r}\!' \, \hat{\cal P}_0 | \psi_1 \rangle
+ {\cal O}(\tau^2) .
\end{eqnarray}

Operating on both sides of this from the left with the ground state projector,
and using the fact that the projector commutes with the Hamiltonian operator
one obtains
\begin{equation} \label{Eq:Ln0S=0}
0
=
 \frac{ |\tau| }{2 } \hat{\cal P}_0 \hat \lambda
 \hat {\overline S}_\mathrm{r}\!' \, | \psi_1 \rangle
- \frac{ |\tau| }{2 } \hat \lambda
 \hat {\overline S}_\mathrm{r}\!' \, \hat{\cal P}_0 | \psi_1 \rangle .
\end{equation}
Since $ \hat {\overline S}_\mathrm{r}\!' \, \hat{\cal P}_0 | \psi_1 \rangle
= 0$,
this proves that
$\hat{\cal P}_0 \hat \lambda
 \hat {\overline S}_\mathrm{r}\!' = 0$.
 Hence in the entropy representation one must have
$\lambda_{0n}^\mathrm{S}=\lambda_{n0}^\mathrm{S}=0$,
since $\hat \lambda $ is a symmetric operator.
This says that the dissipative or drag operator
does not couple the ground and excited states.

Using this, the projector terms cancel
and the expression for the evolution of the full wave function is
finally given by
\begin{eqnarray}
 \left| \overline \psi(t_2|\psi_1,t_1) \right>
& = &
| \psi_1 \rangle
+ \frac{\tau}{i\hbar} \hat {\cal H} | \psi_1 \rangle
+ \frac{ |\tau| }{2 } \hat \lambda
 \hat {\overline S}_\mathrm{r}\!' \, | \psi_1 \rangle
\nonumber \\ &  & \mbox{ }
+ {\cal O}(\tau^2) .
\end{eqnarray}
This is the dissipative Schr\"odinger equation.

The final term in the dissipative Schr\"odinger equation.
is  the gradient of the first entropy.
This provides the thermodynamic driving force
toward the most likely state.
(Actually, this is only true if one has not evolved
into the final equilibrium state.
It is shown below that if the equilibrium
probability operator is stationary,
then the drag operator
must commute with the entropy operator,
and there is no mixing of entropy states due to this term.)
It is the exact analogue of the dissipative term in the classical
Langevin equation.
In that case the dissipation is linearly proportional to the velocity,
which itself is proportional to the velocity gradient of the entropy.

The operator $\hat \lambda$ may be called the dissipative operator,
or statistical drag operator.
From the symmetry requirements it must be Hermitian and real.
Apart from this its magnitude can be chosen within wide bounds.
In so far as the norm of the ground state projection $N(\overline \psi)$
is a constant of the motion (see below),
it can be taken out of the thermodynamic force operator
and used effectively to rescale the dissipative operator.
The thermodynamic force operator reflects the exchange
with the thermal reservoir,
and, as is always the case with the reservoir formalism,
it is an abstraction of reality;
the final results are not sensitive to its precise value
(see next).

\subsection{Stochastic, Dissipative Schr\"odinger Equation}

\subsubsection{Stochastic Contribution}

Since the evolution of the sub-system wave function
is determined in part by the interactions with the reservoir,
and since the wave function of the reservoir is unknown,
there must be a random element to the sub-system evolution,
which is to say that it is only determined in a probabilistic sense,
and it may not be the same each time that the sub-system visits the
same sub-system wave state.
Therefore, one must add a stochastic operator
to the above deterministic equation
to give  the stochastic, dissipative Schr\"odinger equation,
\begin{eqnarray} \label{Eq:SDSE}
 \left| \psi_2 \right>
& = &
\left| \psi_1 \right>
+
\frac{t_{21}}{i\hbar} \hat {\cal H} \left| \psi_1 \right>
+\frac{ |t_{21}| }{2 } \hat \lambda \,  \hat {\overline S}_\mathrm{r}\!'
\left| \psi_1 \right>
+ \hat{\cal R} \left| \psi_1 \right>
\nonumber \\ & \equiv &
\left[ \hat{\mathrm I} + \hat u(t_{21}) + \hat{\cal R}  \right]
\left| \psi_1 \right>
\nonumber \\ & \equiv &
\hat{\cal U}(t_{21}) \left| \psi_1 \right>.
\end{eqnarray}
This is valid to linear order in the time step.
Recall that the propagator is non-linear
since the thermodynamic force
is inversely proportional to the norm of the ground state projection,
$N(\overline \psi_1)$.
However,
in so far as $N(\overline \psi)$ is a constant of the motion (see below),
it can be taken out of the thermodynamic force operator
and used effectively to rescale the dissipative operator.

Note that for this open system and its stochastic dissipative
Schr\"odinger equation,
the time reversed propagator is not equal to the inverse,
$\hat {\cal U}(t_1,t_2) \ne \hat {\cal U}(t_2,t_1)^{-1} $.
This contrasts with an isolated system,
where the explicit form for the adiabatic time propagator,
$\hat U^0(t_2,t_1) \equiv \exp - i (t_2-t_1)\hat {\cal H} /\hbar$,
shows that
$\hat U^0(t_1,t_2) = \hat U^0(t_2,t_1)^{-1} $.
The difference between the two cases in the presence
of the irreversible terms,
those proportional to $|t_{21}|$,
in the  stochastic, dissipative Schr\"odinger equation.

\subsubsection{Microscopic Reversibility}

Let
\begin{equation}
| \psi_2 \rangle = \hat {\cal U}(t_{21}) | \psi_1 \rangle ,
\end{equation}
and
\begin{equation}
| \psi_4 \rangle = \hat {\cal U}(t_{21}) | \psi_3 \rangle .
\end{equation}
These are stochastic equations.
Choose $\psi_3 = \psi_2^*$.
Then from microscopic reversibility one should have, on average,
$\psi_4 = \psi_1^*$, or
\begin{eqnarray}
| \psi_1^* \rangle
& = & \hat {\cal U}(t_{21}) | \psi_2^* \rangle
\nonumber \\ & = &
\hat {\cal U}(t_{21}) \,\hat {\cal U}(t_{21})^* | \psi_1^* \rangle .
\end{eqnarray}
The second equality invokes the complex conjugate of the first expression.
Since these are stochastic equations,
this should hold on average,
which leads to
\begin{equation}
\left< \hat {\cal U}(t_{21}) \,\hat {\cal U}(t_{21})^*  \right>_\mathrm{stoch}
= \hat{\mathrm I} .
\end{equation}

\subsubsection{Unitary Condition}

This is in fact one form of the unitary condition,
which ultimately comes from the reduction condition
on the transition probability.\cite{QSM3}
The unitary condition can be written in eight equivalent forms
using the fact that the identity operator is invariant
under the operations of Hermitian conjugation,
complex conjugation,
and transposition,
\begin{eqnarray}
\hat{\mathrm I}
& = &
\left< \hat {\cal U}(t_{21}) \, \hat {\cal U}(t_{21})^\dag
  \right>_\mathrm{stoch}
=
\left< \hat {\cal U}(t_{21})^\dag \, \hat{\cal U}(t_{21})
  \right>_\mathrm{stoch}
\nonumber \\ & = &
\left< \hat {\cal U}(t_{21})^\mathrm{T} \, \hat {\cal U}(t_{21})^*
  \right>_\mathrm{stoch}
=
\left< \hat {\cal U}(t_{21})^* \, \hat {\cal U}(t_{21})^\mathrm{T}
  \right>_\mathrm{stoch}
\nonumber \\ & = &
\left< \hat {\cal U}(t_{21})^* \, \hat {\cal U}(t_{21})
  \right>_\mathrm{stoch}
=
\left< \hat {\cal U}(t_{21}) \, \hat {\cal U}(t_{21})^*
  \right>_\mathrm{stoch}
\nonumber \\ & = &
\left< \hat {\cal U}(t_{21})^\dag \, \hat {\cal U}(t_{21})^\mathrm{T}
  \right>_\mathrm{stoch}
=
\left< \hat {\cal U}(t_{21})^\mathrm{T} \, \hat {\cal U}(t_{21})^\dag
  \right>_\mathrm{stoch} .
\end{eqnarray}
It can be seen that the fifth and sixth of these correspond to
microscopic reversibility.

The unitary condition
gives an expression for the variance of the stochastic operator.
To linear order in the time step $t_{21}$,
the first form of the unitary condition is
\begin{eqnarray}
\hat{\mathrm I} & = &
\left<
\left[ \hat{\mathrm I} + \hat u(t_{21}) + \hat{\cal R} \right]
\left[ \hat{\mathrm I} + \hat u(t_{21}) + \hat{\cal R} \right]^\dag
\right>_\mathrm{stoch}
\nonumber \\ & = &
\hat{\mathrm I} + \hat u(t_{21})  + \hat u(t_{21})^\dag
+ \left< \hat{\cal R} \hat{\cal R}^\dag
\right>_\mathrm{stoch}
+ {\cal O}(\tau^2)
\\ \nonumber & = &
\hat{\mathrm I}
+
\frac{|t_{21}| }{2}
\left[ \hat \lambda \hat {\overline S}_\mathrm{r}\!'
+ \hat {\overline S}_\mathrm{r}\!'\,\!^\dag \hat \lambda^\dag \right]
+
\left< \hat{\cal R} \hat{\cal R}^\dag\right>_\mathrm{stoch}
+ {\cal O}(t_{21}^2) .
\end{eqnarray}
The final equality follows after the stochastic average
and neglecting terms higher than linear order in $t_{21}$.
Note that $\hat \lambda$ and $\hat {\overline S}_\mathrm{r}\!'$ are Hermitian;
in fact they are real and transpose symmetric.
This leads to eight expressions for the variance,
\begin{eqnarray} 
\lefteqn{
\frac{-|t_{21}| }{2}
\left[ \hat \lambda \hat {\overline S}_\mathrm{r}\!'
+ \hat {\overline S}_\mathrm{r}\!' \hat \lambda\right]
} \nonumber \\
& = &
\left< \hat{\cal R} \, \hat{\cal R}^\dag  \right>_\mathrm{stoch}
=
\left< \hat{\cal R}^\dag \, \hat{\cal R}  \right>_\mathrm{stoch}
=
\left< \hat{\cal R}^\mathrm{T} \, \hat{\cal R}^*  \right>_\mathrm{stoch}
\nonumber \\ & = &
\left< \hat{\cal R}^* \, \hat{\cal R}^\mathrm{T}  \right>_\mathrm{stoch}
=
\left< \hat{\cal R}^* \, \hat{\cal R}  \right>_\mathrm{stoch}
=
\left< \hat{\cal R} \, \hat{\cal R}^*   \right>_\mathrm{stoch}
\nonumber \\ & = &
\left< \hat{\cal R}^\dag \, \hat{\cal R}^\mathrm{T}  \right>_\mathrm{stoch}
=
\left< \hat{\cal R}^\mathrm{T} \, \hat{\cal R}^\dag   \right>_\mathrm{stoch} .
\end{eqnarray}
The simplest way to make these eight expressions equivalent
is to take the stochastic operator
to be real and transpose symmetric,
$\hat{\cal R} = \hat{\cal R}^\mathrm{T} = \hat{\cal R}^*$.
(One could argue that reality reflects the time symmetry of the reservoir.)
In this case  the unitary condition is simply
\begin{equation} \label{Eq:RR-unit}
\left< \hat{\cal R}^2   \right>_\mathrm{stoch}
=
\frac{-|t_{21}| }{2}
\left[ \hat \lambda \hat {\overline S}_\mathrm{r}\!'
+ \hat {\overline S}_\mathrm{r}\!' \hat \lambda\right] .
\end{equation}
Since  $\hat {\overline S}_\mathrm{r}\!'$  is negative semi-definite,
and $\hat \lambda$ is positive semi-definite,
one can see from this that the variance of the stochastic operator
is Hermitian, positive semi-definite,
and proportional to the length of the time step.
As is argued below,
this is the fundamental form
of the fluctuation-dissipation theorem.

Taking the stochastic operator to be
to be real and transpose symmetric
means that it is  Hermitian
$\hat{\cal R}^\dag =\hat{\cal R} $.
As just mentioned it depends upon the absolute value of the times step.
It will be shown below that in the case that the probability operator
is stationary, the dissipative operator
and the thermodynamic force operator commute
$\hat \lambda \hat{\overline S}_\mathrm{r}\!'
= \hat{\overline S}_\mathrm{r}\!' \hat \lambda$.
Both are also Hermitian operators.
In this case,
from the explicit expression for the stochastic dissipative time propagator,
Eq.~(\ref{Eq:SDSE}),
one sees therefore that the time propagator has the symmetry
\begin{equation}
\hat{\cal U}(t_{21})^\dag = \hat{\cal U}(t_{12}).
\end{equation}
This is equivalent to the symmetry identified
for the classical conditional transition probability density
in Eq.~(7.164) of Ref.~\onlinecite{NETDSM},
$\wp({\bf \Gamma}_2 | {\bf \Gamma}_1,\Delta_t)
= \wp({\bf \Gamma}_2^\dag | {\bf \Gamma}_1^\dag,-\Delta_t)$.

\subsubsection{Stationarity Condition}

The time evolution
of the probability operator is\cite{QSM3}
\begin{eqnarray}
\hat \wp(t)
& = &
\left< \hat {\cal U}(t) \, \hat \wp(0) \, \hat{\cal U}(t)^\dag
\right>_\mathrm{stoch}
\nonumber \\ & = &
\left< \hat {\cal U}(t)^\dag \, \hat \wp(0) \, \hat{\cal U}(t)
\right>_\mathrm{stoch}.
\end{eqnarray}

In so far as the probability operator
equals the density operator,
this is like a stochastic form of the conventional Lindblad master equation
in the Krauss representation.
It should be noted that the present formula,
in addition to being stochastic,
has been derived with explicit terms that abstract from a specific model
for the reservoir,
and that it obeys
the exact unitary and irreversibility symmetry rules
that are derived from the underlying Schr\"odinger equation
for the total isolated system and the equilibrium nature of the reservoir.

Under the equilibrium stochastic, dissipative equations of motion,
the Maxwell-Boltzmann probability operator should be stationary,
\begin{eqnarray} \label{Eq:Stat-Cond-gen}
\hat \wp_\mathrm{MB}
& = &
\left< \hat {\cal U}(t) \, \hat \wp_\mathrm{MB} \, \hat{\cal U}(t)^\dag
\right>_\mathrm{stoch}
\nonumber \\ & = &
\left< \hat {\cal U}(t)^\dag \, \hat \wp_\mathrm{MB} \, \hat{\cal U}(t)
\right>_\mathrm{stoch} ,
\end{eqnarray}
where
$\hat \wp_\mathrm{MB}
= Z(T)^{-1} e^{\hat S_\mathrm{r} /k_\mathrm{B}}
= Z(T)^{-1} e^{-\hat {\cal H} /k_\mathrm{B}T}$.

The stationarity condition is a very strong condition to impose.
It will be shown below that
it implies that there is no mixing between entropy modes:
not only is the Maxwell-Boltzmann probability operator stationary,
but any probability operator that corresponds
to the density matrix constructed from the initial wave function
is constant on the trajectory;
even a non-equilibrium distribution of entropy modes is conserved.
In a literal sense this is consistent with  the present theory
being purely an equilibrium theory:
neither the equilibrium nor any other distribution of entropy states evolves.
The theory, however, would be somewhat more useful in a practical sense
if the stochastic, dissipative propagator
evolved an initial non-equilibrium distribution
over time into the equilibrium distribution,
so that the density matrix constructed from the wave function
on the trajectory converged to the equilibrium probability operator.
This can only be done if the propagator is constructed
to violate the stationarity condition.
For the present the consequences of imposing the stationarity condition
will be pursued,
and further discussion of the issue is deferred.

To linear order in the time step,
the two forms of the stationarity condition yield
\begin{equation}
\frac{-|t_{21}| }{2}
\left[ e^{\hat S_\mathrm{r} /k_\mathrm{B} }
\hat \lambda \hat {\overline S}_\mathrm{r}\!'
 + \hat {\overline S}_\mathrm{r}\!' \hat \lambda
e^{\hat S_\mathrm{r} /k_\mathrm{B} } \right]
 =
\left< \hat{\cal R} \,
e^{\hat S_\mathrm{r} /k_\mathrm{B} }
\,  \hat{\cal R}^\dag  \right>_\mathrm{stoch} ,
\end{equation}
and
\begin{equation}
\frac{-|t_{21}| }{2}
\left[ e^{\hat S_\mathrm{r} /k_\mathrm{B} }
\hat {\overline S}_\mathrm{r}\!' \hat \lambda
 +  \hat \lambda \hat {\overline S}_\mathrm{r}\!'
e^{\hat S_\mathrm{r} /k_\mathrm{B} } \right]
=
\left< \hat{\cal R}^\dag \,
e^{\hat S_\mathrm{r} /k_\mathrm{B} }
\,  \hat{\cal R}  \right>_\mathrm{stoch} .
\end{equation}
The adiabatic evolution has canceled here because
the Hamiltonian operator commutes with the probability operator.
Since the stochastic operator is taken to be real and symmetric,
the two right hand sides are equal.

If  $\hat{\cal R}$ and $\hat S_\mathrm{r}$
(equivalently $\hat  \wp$)
commute,
then this is automatically satisfied
if the unitary condition is satisfied.
This will occur if $\hat{\cal R}$ is constructed
from entropy eigenfunctions,
which is indeed the case as is discussed below.

\subsubsection{Entropy Representation in the Stationary Case}

One can obtain the explicit form of the dissipative and stochastic operators
by invoking the entropy representation in the stationary case.
For the reasons discussed below,
in practical terms the stationary condition is too restrictive.
For computational approaches to equilibrium systems
one should only impose the unitary condition,
in which case
most of the results given in this section do not hold.

In order to simplify the notation,
the entropy eigenfunctions will be written with a single Roman letter,
$ \hat { S}_\mathrm{r}\, | \zeta^S_n \rangle
=  S_{n}   | \zeta^S_n \rangle$.
A more precise notation would single out the principle
and degenerate entropy states,
$ \hat { S}_\mathrm{r}\, | \zeta^S_{\alpha g} \rangle
=  S_{\alpha}   | \zeta^S_{\alpha g} \rangle$.
The sums that follow are over all entropy states;
references to the vanishing of correlations between states,
or to only intra-state terms surviving should be taken to mean
principle entropy states only.

The entropy eigenfunctions
are also eigenfunctions of the thermodynamic force operator,
$ \hat {\overline S}_\mathrm{r}\!' \, | \zeta^S_n \rangle
= \overline S_{n}\!\!'  | \zeta^S_n \rangle$,
with the eigenvalue being
\begin{equation}
\overline S_{n}\!\!'  \equiv
\frac{k_\mathrm{B} }{N(\overline \psi)}
\left[ \frac{ e^{S_n/k_\mathrm{B}} }{ e^{S_0/k_\mathrm{B}} }
- 1 \right].
\end{equation}
These are obviously negative (or zero for the ground state),
and the depend upon the projection of wave function onto the ground state.



In the entropy representation the unitary condition,
Eq.~(\ref{Eq:RR-unit}), is
\begin{equation} \label{Eq:unit-ent-rep}
\sum_l
\left< R_{ml}^{\mathrm{S}} R_{ln}^\mathrm{S} \right>_\mathrm{stoch}
=
\frac{-|t_{21}| }{2}
\left[ \lambda_{mn}^\mathrm{S}  \overline S_{n}\!\!' \,
 + \overline S_{m}\!\!' \, \lambda_{mn}^\mathrm{S} \right].
\end{equation}
Recall that $\hat {\cal R}$ has been taken to be
a real, symmetric operator.
One stationarity condition is
\begin{eqnarray}
\lefteqn{
 \sum_l e^{ S_l /k_\mathrm{B} }
\left< R_{ml}^{\mathrm{S}} R_{ln}^\mathrm{S} \right>_\mathrm{stoch}
} \nonumber \\
& = &
\frac{-|t_{21}| }{2}
\left[
e^{ S_m /k_\mathrm{B} } \lambda_{mn}^\mathrm{S}  \overline S_{n}\!\!' \,
 + \overline S_{m}\!\!' \, \lambda_{mn}^\mathrm{S} e^{ S_n /k_\mathrm{B} }
\right] ,
\end{eqnarray}
and the other is
\begin{eqnarray}
\lefteqn{
 \sum_l e^{ S_l /k_\mathrm{B} }
\left< R_{ml}^{\mathrm{S}} R_{ln}^\mathrm{S} \right>_\mathrm{stoch}
} \nonumber \\
& = &
\frac{-|t_{21}| }{2}
\left[
e^{ S_m /k_\mathrm{B} }  \overline S_{m}\!\!' \,\lambda_{mn}^\mathrm{S}
 + \lambda_{mn}^\mathrm{S} \overline S_{n}\!\!' \, e^{ S_n /k_\mathrm{B} }
\right].
\end{eqnarray}

Subtracting these last two,
one obtains
\begin{eqnarray}
0 & = &
e^{ S_m /k_\mathrm{B} } \lambda_{mn}^\mathrm{S}  \overline S_{n}\!\!' \,
 + \overline S_{m}\!\!' \, \lambda_{mn}^\mathrm{S} e^{ S_n /k_\mathrm{B} }
\nonumber \\ && \mbox{ }
-
e^{ S_m /k_\mathrm{B} }  \overline S_{m}\!\!' \,\lambda_{mn}^\mathrm{S}
- \lambda_{mn}^\mathrm{S} \overline S_{n}\!\!' \, e^{ S_n /k_\mathrm{B} }
\nonumber \\ &=&
e^{ S_m /k_\mathrm{B} } \lambda_{mn}^\mathrm{S}
\left[ \overline S_{n}\!\!' \, -  \overline S_{m}\!\!'\, \right]
\nonumber \\ && \mbox{ }
+ \lambda_{mn}^\mathrm{S} e^{ S_n /k_\mathrm{B} }
\left[  \overline S_{m}\!\!' \, - \overline S_{n}\!\!' \right]
\nonumber \\ &=&
-\lambda_{mn}^\mathrm{S}
\left[  \overline S_{m}\!\!' \, - \overline S_{n}\!\!' \right]
\left[ e^{ S_m /k_\mathrm{B} } - e^{ S_n /k_\mathrm{B} }\right] .
\end{eqnarray}
Since $S_m > S_n$ if $m > n$
(more precisely, $S_{\alpha g} > S_{\gamma h}$ if $ \alpha > \gamma$),
this implies
\begin{equation}
\lambda_{mn}^\mathrm{S}
=
\lambda_{nn}^\mathrm{S} \delta_{mn}.
\end{equation}
Hence the drag or dissipative operator must be diagonal
in the entropy representation
and so it must be constructed from the entropy eigenfunctions,
\begin{equation}
\hat \lambda \equiv
\sum_n  \lambda_n  \left| \zeta^S_n \right>  \left< \zeta^S_n \right|.
\end{equation}
(The diagonalization holds for the principle entropy states only;
the most general dissipative operator is of the form
$ \hat\lambda \equiv
\sum_{\alpha, gh}  \lambda_{\alpha, gh}
\left| \zeta^S_{\alpha g} \right> \left< \zeta^S_{\alpha h} \right|$.)
This gives
$\hat \lambda | \zeta^S_n \rangle = \lambda_n | \zeta^S_n \rangle$.
Because $\hat \lambda $ is a real Hermitian operator,
the $\lambda_n$ must be real.
It may be called the drag coefficient for the state $ \zeta^S_n $.
One can set $\lambda_0 = 0$ since it always multiplies
the thermodynamic force, which vanishes  in the ground state,
$\overline S_{0}\!\!' = 0$.
(This results holds even when the stationarity condition doesn't.)
Obviously,
this construction means that
the drag operator  commutes with the entropy operator,
the thermodynamic force operator, the probability operator,
and the Hamiltonian  operator.
One sees from this that there can be no dissipative coupling
between principle entropy states.
There can be dissipative coupling between degenerate entropy states.

It should be noted that the dissipative operator
is only diagonal if the stationarity condition holds.
If one does not insist upon the stationarity condition,
then there can be dissipative coupling between the principle entropy states.
In this case the variance of the stochastic operator
is determined by the unitary condition alone, Eq.~(\ref{Eq:unit-ent-rep}).
In this case one expects that the thermodynamic force
will tend to drive the wave function toward high entropy states,
and that in combination with the stochastic force
it will  cause the density matrix to converge on
the equilibrium probability operator.

With this diagonal form for the dissipative operator
the stationarity condition becomes
\begin{equation}
\sum_l
\left< R_{ml}^{\mathrm{S}} R_{ln}^\mathrm{S} \right>_\mathrm{stoch}
=
-|t_{21}|
\lambda_{mm}^\mathrm{S}  \overline S_{m}\!\!' \, \delta_{mn} ,
\end{equation}
and either unitary condition is
\begin{equation}
 \sum_l e^{ S_l /k_\mathrm{B} }
\left< R_{ml}^{\mathrm{S}} R_{ln}^\mathrm{S} \right>_\mathrm{stoch}
=
-|t_{21}|
e^{ S_m /k_\mathrm{B} } \lambda_{mm}^\mathrm{S}  \overline S_{m}\!\!' \,
 \delta_{mn}.
\end{equation}

If one sets $n=m=0$, both right hand sides vanish,
since $ \overline S_{m}\!\!'  = 0$.
Since both left hand sides are sums of non-negative terms,
one sees that one must have
\begin{equation}
\langle  R_{0l}^{\mathrm{S}} R_{l0}^\mathrm{S}\rangle_\mathrm{stoch}
= 0 , \;\; l = 0, 1, \ldots
\end{equation}
This means there is no stochastic coupling between the ground state
and the excited states.
It was already shown that there is no dissipative coupling in the
ground state.
There is no adiabatic coupling
between the ground state and the excited states
(because the Hamiltonian operator commutes with the entropy operator
it is diagonal in the entropy representation).
These mean that in general the ground state evolves independently
of the excited states.

A sufficient condition that makes the left hand sides
of the unitary and stationarity conditions
proportional to a Kronecker-delta function is that
the stochastic operators between different pairs of modes are uncorrelated,
\begin{equation} \label{Eq:RmlRln=dnm}
\left< R_{ml}^{\mathrm{S}} R_{ln}^\mathrm{S} \right>_\mathrm{stoch}
\equiv
\left< R_{ml}^{\mathrm{S}} R_{lm}^\mathrm{S} \right>_\mathrm{stoch}
 \delta_{mn}.
\end{equation}
The conditions now become
\begin{equation}
\sum_{l>0}
\left< R_{ml}^{\mathrm{S}} R_{lm}^\mathrm{S} \right>_\mathrm{stoch}
=
-|t_{21}|
\lambda_{mm}^\mathrm{S}  \overline S_{m}\!\!' \, ,
\end{equation}
and
\begin{equation}
 \sum_{l>0} e^{ S_l /k_\mathrm{B} }
\left< R_{ml}^{\mathrm{S}} R_{lm}^\mathrm{S} \right>_\mathrm{stoch}
=
-|t_{21}|
e^{ S_m /k_\mathrm{B} } \lambda_{mm}^\mathrm{S}  \overline S_{m}\!\!' \, .
\end{equation}

If one  multiples both sides of the first by
$ e^{ S_m /k_\mathrm{B} }$, and subtracts these two,
then one obtains
\begin{equation}
 \sum_{l>0} \left[ e^{ S_m /k_\mathrm{B} } - e^{ S_l /k_\mathrm{B} } \right]
\left< R_{ml}^{\mathrm{S}} R_{lm}^\mathrm{S} \right>_\mathrm{stoch}
=0 .
\end{equation}
Setting $m=1$, one sees that both factors in the summand
are non-negative, since $S_m > S_n$ if $m > n$.
Hence one concludes
that there is no stochastic coupling between the first excited state
and the more excited states,
$\left< R_{1l}^{\mathrm{S}} R_{l1}^\mathrm{S} \right>_\mathrm{stoch}
=0$ if $l>1$.
The equation now becomes
\begin{equation}
 \sum_{l>1} \left[ e^{ S_m /k_\mathrm{B} } - e^{ S_l /k_\mathrm{B} } \right]
\left< R_{ml}^{\mathrm{S}} R_{lm}^\mathrm{S} \right>_\mathrm{stoch}
=0 .
\end{equation}
Setting $m=2$, one sees that there is no stochastic coupling
to the second excited state from the even more excited states.
Continuing in this fashion,
one concludes that each principle entropy state
is only coupled stochastically to itself,
\begin{equation}
\left< R_{ml}^{\mathrm{S}} R_{lm}^\mathrm{S} \right>_\mathrm{stoch}
=
\left< R_{mm}^{\mathrm{S}} R_{mm}^\mathrm{S} \right>_\mathrm{stoch}
\delta_{ml} .
\end{equation}

It follows that the stochastic operator, like the dissipative operator,
is diagonal in the entropy representation,
\begin{equation}
\hat{\cal R} \equiv
\sum_n r_n \left| \zeta^S_n \right> \left< \zeta^S_n \right| .
\end{equation}
(Again, the diagonalization holds for the principle entropy states only;
the most general stochastic operator is of the form
$ \hat{\cal R} \equiv
\sum_{\alpha, gh}  r_{\alpha, gh}
\left| \zeta^S_{\alpha g} \right> \left< \zeta^S_{\alpha h} \right|$.)
Here the  $r_n$ are random real  numbers of zero mean.
With this $\hat {\cal R}   | \zeta^S_n \rangle
= r_n   | \zeta^S_n \rangle $,
and the stochastic operator commutes with the entropy operator.
This means that the unitary condition automatically
implies the stationarity condition,
and that the variance of the coefficients is given by
\begin{eqnarray} \label{Eq:FD-thm}
\left<  r_n^2 \right>_\mathrm{stoch}
& = &
-|t_{21}| \overline S_{n}\!\!'  \lambda_n .
\end{eqnarray}
Note that the right hand side is inversely proportional
to the magnitude of the projection of the current wave function
onto the ground state.
(Actually, because there is no mixing with the ground state,
the norm of the ground state projection is a constant of the motion
that can be taken out of the thermodynamic force operator
and incorporated into a renormalized dissipative force operator.)
This is  a form of the quantum fluctuation-dissipation theorem.
Since the thermodynamic force eigenfunctions are negative or zero,
and since the drag operator coefficients are positive or zero,
one can see that the variance of the random  coefficients
is positive or zero.

As was the case with the dissipative operator,
the stochastic operator is only diagonal if the stationarity condition holds.
If one does not insist upon the stationarity condition,
then there can be stochastic coupling between the principle entropy states,
 Eq.~(\ref{Eq:unit-ent-rep}).
In this case one expects that combination
of the thermodynamic force and the stochastic force
it will  cause the density matrix to converge to
the equilibrium probability operator.

The simplest implementation of the above
is to use a single dissipative coefficient
and a single stochastic coefficient.
Both  operators can be made proportional to
the thermodynamic force operator,
\begin{equation}
\hat \lambda = - \lambda \hat S_\mathrm{r}'(\overline \psi)
,\mbox{ and }
\hat R =  r \hat S_\mathrm{r}'(\overline \psi) ,
\end{equation}
with $\lambda$ real and positive.
In this case also the unitary condition and the stationary condition
yield the same variance, namely
\begin{equation} \label{Eq:FD-thm-2}
\left< r^2  \right>_\mathrm{stoch}
=
|\tau|  \lambda .
\end{equation}

Equation (\ref{Eq:FD-thm}) and  Eq.~(\ref{Eq:FD-thm-2})
are forms of the quantum fluctuation dissipation theorem.
They say that the variance of the stochastic operator,
which controls the fluctuation,
must be linearly proportional to the drag operator,
which controls the dissipation.
The variance is proportional to the duration of the time step,
so it is an irreversible contribution
to the evolution of the wave function.

The stationarity condition is an optional condition
that should only be imposed if the initial wave function
obeys the equilibrium distribution.
It is only if this condition is imposed
that the dissipative and stochastic operators are diagonal
in the entropy basis and that the entropy modes do not mix.
In practical terms it is better \emph{not} to impose
the stationarity condition as this allows an arbitrary starting wave function.
One should impose the unitary condition,
with the dissipative operator and the stochastic operator
\emph{not} being diagonal in the entropy basis.
In this way the trajectory of an arbitrary starting wave function
will yield a density matrix that converges to the equilibrium
probability operator and fluctuate about it,
with the superposition states averaging to zero
over the trajectory.
A time average of the expectation value of an operator
will then equal the equilibrium statistical average.
The unitary condition must be imposed,
and when it is the only such condition on the propagator
the operators are not diagonal
and do not need to be represented in the entropy basis.
For this reason one can arguably regard the unitary condition
as \emph{the} quantum fluctuation dissipation theorem.

\subsection{Transition Probability Operator
and Time Correlation Function}

\subsubsection{Time Correlation Function}

The transition probability operator is
just the exponential of the second entropy operator,
\begin{equation}
\hat \wp^{(2)}(\tau,T)
=
\frac{1}{Z^{(2)}(\tau,T)}
e^{ \hat S^{(2)}(\tau,T) /k_\mathrm{B} } .
\end{equation}
With it, the statistical average of a two-time operator
$\hat O^{(2)}$ is
\begin{eqnarray}
\left< \hat O^{(2)} \right>_\mathrm{\tau,T}
& = &
\int \mathrm{d} \psi_1 \,  \mathrm{d} \psi_2 \;
\frac{
\left< \psi_2 ,  \psi_1 \right|
\hat \wp^{(2)} \hat O^{(2)}
\left| \psi_1 ,  \psi_2 \right>
}{ \left<  \psi_2 ,  \psi_1 | \psi_1 ,  \psi_2 \right> }
\nonumber \\ & = &
\frac{1}{Z^{(2)}}
\int \mathrm{d} \underline \psi_1 \, \mathrm{d}  \underline \psi_2 \;
\frac{1}{ N(\psi_1) N(\psi_2) }
\nonumber \\ &  & \mbox{ } \times
\sum_{\stackrel{\scriptstyle m_2,n_2}{m_1,n_1}}
\psi_{2,m_2}^* \psi_{2,n_2} \psi_{1,m_1}^* \psi_{1,n_1}
\nonumber \\ &  & \mbox{ } \times
\left\{ e^{ \hat S^{(2)}(\tau,T) /k_\mathrm{B} } \hat O^{(2)}
\right\}_{\stackrel{\scriptstyle m_2,n_2}{m_1,n_1}}
\nonumber \\ & = &
\frac{\mbox{const.}}{Z^{(2)}} \sum_{m_2,m_1}
\left\{ e^{ \hat S^{(2)}(\tau,T) /k_\mathrm{B} } \hat O^{(2)}
\right\}_{\stackrel{\scriptstyle m_2,m_2}{m_1,m_1}}
\nonumber \\ & = &
\mathrm{Tr}^{(2)}
\left\{   \hat \wp^{(2)}(\tau,T)  \hat O^{(2)} \right\} .
\end{eqnarray}
In passing to the third equality,
the same trick as in Eq.~(\ref{Eq:O-vN2}) has been used,
namely that all the terms in the integrand are odd
except those with $m_2 = n_2$ and $m_1 = n_1$.
This is the dual collapse of the wave functions
at the terminii of the transition.
Once the representation of the product of the operators
has been taken outside of the integral,
what remains is the same for all indeces,
and hence the integral is a constant that can be taken outside
of the sum and incorporated into the partition function.

The transition probability operator is
$ \hat \wp^{(2)}(\tau,T) \equiv
e^{ \hat S^{(2)}(\tau,T) /k_\mathrm{B} } /Z'^{(2)}$
with the partition function being
\begin{eqnarray}
Z'^{(2)}(\tau,T)
& = &
\mathrm{Tr}^{(2)} e^{ \hat S^{(2)}(\tau,T) /k_\mathrm{B} }
\\ \nonumber & = &
\sum_{m_2,m_1}
\left< \zeta_{m_2} , \zeta_{m_1} \right|
e^{ \hat S^{(2)}(\tau,T) /k_\mathrm{B} }
\left| \zeta_{m_1} , \zeta_{m_2} \right> ,
\end{eqnarray}
with the $\zeta$ being an arbitrary orthonormal basis.
The elements of the transition matrix are explicitly
\begin{equation}
 \wp^{(2)}_{\stackrel{\scriptstyle m_2,n_2}{m_1,n_1}}
 =
 \left< \zeta_{m_2} , \zeta_{m_1} \right|
\hat \wp^{(2)}(\tau,T)
\left| \zeta_{n_1} , \zeta_{n_2} \right> ,
\end{equation}
and similarly for the operator matrix.
Hence the two-time trace is explicitly
\begin{equation}
\mathrm{Tr}^{(2)}
\left\{   \hat \wp^{(2)}  \hat O^{(2)} \right\}
=
\sum_{\stackrel{\scriptstyle m_2,n_2}{m_1,n_1}}
 \wp^{(2)}_{\stackrel{\scriptstyle m_2,n_2}{m_1,n_1}}
O^{(2)}_{\stackrel{\scriptstyle n_2,m_2}{n_1,m_1}} .
\end{equation}

A common quantity is the time correlation of two one-time operators.
To obtain this one can introduce the diagonal two-time operator,
$ \hat D^{(2)}_{BA}$,
which has the expectation
\begin{equation}
\frac{
\left< \psi_2 ,  \psi_1 \right|
\hat D^{(2)}_{BA}
\left| \psi_1 ,  \psi_2 \right>
}{N(\psi_1) N(\psi_2)}
=
\frac{\left< \psi_2 \right| \hat B \left| \psi_2 \right>}{N(\psi_2)}
\frac{\left< \psi_1 \right| \hat A \left| \psi_1 \right>}{N(\psi_1)} .
\end{equation}
The time correlation function
will shortly be expressed in terms of this and
the transition probability operator.


Choose $\langle \hat A \rangle_\mathrm{stat}
= \langle \hat B \rangle_\mathrm{stat} = 0$.
Define the time correlation function as
\begin{equation}
C_{BA}(\tau)
= \left< \hat B(\tau) \hat A(0)  \right>_\mathrm{stat}
= \left< \hat A(0) \hat B(\tau)  \right>_\mathrm{stat} .
\end{equation}

For a two-time average such as this,
the positional order of the operators is irrelevant because the sign
of $\tau$ gives the order of their application.
This contrasts with a one-time average,
where it is conventional that the position of the operators
in the equation designates the time order of their application,
which is to say that they are applied in order from right to left.
The order of course is significant if the operators don't commute.

In view of the definitions of the two-time and one-time averages,
in the zero time limit one has
\begin{equation}
C_{BA}(0^+) = \left< \hat B \hat A  \right>_\mathrm{stat}
\mbox{ and }
C_{BA}(0^-) = \left< \hat A \hat B  \right>_\mathrm{stat} .
\end{equation}
The right hand sides are one-time averages,
which are here signified by the absence of a time argument.
These expressions mean that  if the operators don't commute,
then there is a discontinuity at $\tau = 0$.

Time homogeneity means that the time correlation function
must be invariant to a shift in the time origin.
Hence first changing $\tau \Rightarrow t+\tau$,
and then setting $t=-\tau$,
such that $\hat B(\tau) \hat A(0) \Rightarrow
\hat B(t+\tau) \hat A(t) \Rightarrow
\hat B(0) \hat A(-\tau) $,
must leave the time correlation function unchanged.
From this one concludes that
\begin{equation} \label{Eq:Ct-stat-sym}
C_{AB}(\tau) = C_{BA}(-\tau).
\end{equation}
This can be called time homogeneity or statistical symmetry.

In view of the above,
the small time expansion of the time correlation function
must be of the form
\begin{equation}
C_{AB}(\tau) =
C_{AB;0} + \widehat \tau C_{AB;0}'
+  |\tau| C_{AB;1} +  \tau C_{AB;1}' ,
\end{equation}
with
\begin{equation}
C_{AB;0} =
\frac{1}{2} \left< \hat A \hat B + \hat B \hat A \right>_\mathrm{stat} ,
\end{equation}
and
\begin{equation}
C_{AB;0}' =
\frac{1}{2} \left< \hat A \hat B - \hat B \hat A \right>_\mathrm{stat} .
\end{equation}

In terms of the diagonal two-time operator
and the transition probability operator,
and not explicitly showing the temperature dependence,
the time correlation function is
\begin{eqnarray}
C_{BA}(\tau) \label{Eq:CBA(tau)}
& \equiv &
\left< \hat D^{(2)}_{BA} \right>_\mathrm{stat}
\nonumber \\ & = &
\mathrm{Tr}^{(2)}
\left\{ \hat \wp^{(2)}(\tau) \hat D^{(2)}_{BA} \right\}
\nonumber \\ & = &
\sum_{\stackrel{\scriptstyle m_2,n_2}{m_1,n_1}}
 \wp^{(2)}_{\stackrel{\scriptstyle m_2,n_2}{m_1,n_1}}(\tau)
 B_{n_2,m_2}  A_{n_1,m_1} .
\end{eqnarray}
If one uses the basis $\{ \zeta_n^A \}$ for $\psi_1$
and $\{ \zeta_n^B \}$ for $\psi_2$,
then the operator matrices are diagonal and this becomes
\begin{eqnarray}
C_{BA}(\tau)
& = &
\sum_{\stackrel{\scriptstyle n_2}{n_1}}
\wp^{(2),BA}_{\stackrel{\scriptstyle n_2,n_2}{n_1,n_1}}(\tau)
B_{n_2,n_2}^B  A_{n_1,n_1}^A
\nonumber \\ & \equiv &
\sum_{n_2,n_1}
\wp^{(2),BA}_{n_2,n_1}(\tau) B_{n_2,n_2}^B  A_{n_1,n_1}^A  .
\end{eqnarray}
One can therefore identify
$\wp^{(2),BA}_{mn}(\tau)$
as the unconditional probability of the transition
between states of the two operators,
$n^A \stackrel{\tau}{\rightarrow } m^B$.
Explicitly this is
\begin{equation}
\wp^{(2),BA}_{mn}(\tau)
\equiv
\wp^{(2),BA}_{\stackrel{\scriptstyle m,m}{n,n}}(\tau)
 =
 \left< \zeta_{m}^B , \zeta_{n}^A \right|
\hat \wp^{(2)}(\tau)
\left| \zeta_{n}^A , \zeta_{m}^B \right> .
\end{equation}

In view of the time homogeneity (statistical) symmetry,
Eq.~(\ref{Eq:Ct-stat-sym}),
the  above representation of the time correlation function,
Eq.~(\ref{Eq:CBA(tau)}),
shows that
\begin{equation} \label{Eq:wp2-stat-sym}
\wp^{(2)}_{\stackrel{\scriptstyle m_2,n_2}{m_1,n_1}}(\tau)
=
\wp^{(2)}_{\stackrel{\scriptstyle m_1,n_1}{m_2,n_2}}(-\tau)  .
\end{equation}
One can see explicitly that this condition is guaranteed by
the statistical symmetry of the second entropy,
Eq.~(\ref{Eq:S2-stat-sym}),
$S^{(2)}( \psi_2, \psi_1;\tau)  = S^{(2)}( \psi_1, \psi_2;-\tau)$,
or, equivalently,  $\hat a(-\tau) = \hat c(\tau)$
and $\hat b(-\tau) = \hat b(\tau)^\dag$.

\subsubsection{Time Correlation with Propagator}

It is desired to write the time correlation function as
a one time trace involving the stochastic time propagator.
The most straightforward expression is
\begin{eqnarray} 
C_{BA}(\tau) & = &
\mbox{Tr}^{(1)} \left< \hat {\cal U}(\tau)^\dag
\hat B  \hat {\cal U}(\tau) \hat A
\hat \wp \right>_\mathrm{stoch}
\nonumber \\ & = &
\mbox{Tr}^{(1)} \left< \hat {\cal U}(-\tau)^\dag
\hat A  \hat {\cal U}(-\tau) \hat B
\hat \wp \right>_\mathrm{stoch} .
\end{eqnarray}
For non-zero $\tau > 0$, both forms unambiguously signify
that the operator $\hat A$ is applied before the operator $\hat B$,
and \emph{vice versa} for  $\tau < 0$.
Unfortunately, due to the discontinuity in the time correlation function,
there is an ambiguity in this expression at $\tau = 0$.

To circumvent this problem,
and because of the convention that
the operators are applied in order from right to left for one-time averages,
one needs to define the  operator
\begin{eqnarray}
\hat \eta_{BA}(\tau)
& = &
\frac{1+\widehat \tau}{2} \hat B
+
\frac{1-\widehat \tau}{2}  \hat A
\nonumber \\ & = &
\left\{ \begin{array}{cc}
\hat B, & \tau > 0 ,\\
\hat A   , & \tau < 0 .
\end{array} \right.
\end{eqnarray}
Recall that $\widehat \tau \equiv \mbox{sign } \tau$.
With this the time correlation function in singlet form is
\begin{equation} \label{Eq:TCF-prop-1}
C_{BA}(\tau) =
\mbox{Tr}^{(1)}
\left< \hat {\cal U}(|\tau|)^\dag
\hat \eta_{BA}(\tau)  \hat {\cal U}(|\tau|) \hat \eta_{BA}(-\tau)
\hat \wp \right>_\mathrm{stoch} .
\end{equation}
The definitions have the effect that the operator
that is applied first in time always occupies the right hand position,
which is the usual convention for non-commuting operators.
In addition, the time propagator always
proceeds in the positive time direction.
Since $\hat \eta_{AB}(-\tau) = \hat \eta_{BA}(\tau) $,
it is clear that  $ C_{BA}(-\tau) =  C_{AB}(\tau)$, as required.

Explicitly the time correlation function is
\begin{eqnarray} \label{Eq:TCF-prop}
C_{BA}(\tau) & = &
\left\{ \begin{array}{ll}
\mbox{Tr}^{(1)} \left< \hat {\cal U}(\tau)^\dag
\hat B  \hat {\cal U}(\tau) \hat A
\hat \wp \right>_\mathrm{stoch} , & \tau > 0 ,\\
\mbox{Tr}^{(1)} \left< \hat {\cal U}(-\tau)^\dag
\hat A  \hat {\cal U}(-\tau) \hat B
\hat \wp \right>_\mathrm{stoch} , & \tau < 0
\end{array} \right.
\nonumber \\ & = &
\left\{ \begin{array}{ll}
\mbox{Tr}^{(1)} \left< \hat {\cal U}(\tau)^\dag
\hat B  \hat {\cal U}(\tau) \hat A
\hat \wp \right>_\mathrm{stoch} , & \tau > 0 ,\\
\mbox{Tr}^{(1)} \left< \hat {\cal U}(\tau)^\dag \hat B \hat {\cal U}(\tau)
\hat \wp \hat A \right>_\mathrm{stoch} , & \tau < 0 .
\end{array} \right.
\end{eqnarray}
The second equality uses the facts that the trace is invariant to cyclic
permutations of the operators,
that $ \hat {\cal U}(-\tau) = \hat {\cal U}(\tau)^\dag$,
and that the equilibrium probability operator and the time propagator commute.
In the limit $\tau \rightarrow 0$,
$\hat {\cal U}(\tau) \rightarrow \hat{\mathrm I}$,
and
\begin{equation}
C_{BA}(0^+) =
\mbox{Tr}^{(1)} \hat B \hat A \hat \wp
=
\left< \hat B \hat A \right>_\mathrm{stat}
\end{equation}
and
\begin{equation}
C_{BA}(0^-) =
\mbox{Tr}^{(1)}\hat A \hat B \hat \wp
=
\left< \hat A \hat B  \right>_\mathrm{stat} ,
\end{equation}
as required.

In a particular representation of the operators,
the average that is the time correlation function for $\tau > 0 $ is
\begin{eqnarray}
C_{BA}(\tau) & = &
\mbox{Tr}^{(1)} \left< \hat {\cal U}(\tau)^\dag
\hat B \hat {\cal U}(\tau) \hat A \hat \wp
\right>_\mathrm{stoch}
, \;\; \tau > 0
\nonumber \\ & = &
\sum_{\stackrel{\scriptstyle m_2,n_2}{m_1,n_1}} \sum_{l}
\left<
{\cal U}_{m_1,n_2}(\tau)^\dag
{\cal U}_{m_2,l}(\tau) \wp_{n_1,m_1} \right>_\mathrm{stoch}
\nonumber \\ & & \mbox{ } \times
B_{n_2,m_2} A_{l,n_1}
\nonumber \\ & = &
\mbox{Tr}^{(1)} \left< \hat {\cal U}(\tau)^\dag
\hat \wp \hat B \hat {\cal U}(\tau) \hat A
\right>_\mathrm{stoch}
\nonumber \\ & = &
\sum_{\stackrel{\scriptstyle m_2,n_2}{m_1,n_1}} \sum_{l}
\left<
{\cal U}_{m_1,l}(\tau)^\dag
{\cal U}_{n_2,n_1}(\tau) \wp_{l,m_2} \right>_\mathrm{stoch}
\nonumber \\ & & \mbox{ } \times
B_{m_2,n_2} A_{n_1,m_1} .
\end{eqnarray}
The third equality uses the fact that
the time propagator and the probability operator commute.
Comparing this to the third equality in Eq.~(\ref{Eq:TCF-prop-1}),
this implies that the representation of the transition
probability operator in terms of time propagators is
\begin{eqnarray}
\lefteqn{
\wp^{(2)}_{\stackrel{\scriptstyle m_2,n_2}{m_1,n_1}}(\tau)
} \\ \nonumber & = &
\sum_{l}
\left<
{\cal U}_{l,n_2}(\tau)^\dag
{\cal U}_{m_2,n_1}(\tau) \wp_{m_1,l} \right>_\mathrm{stoch}
, \;\; \tau > 0
\nonumber \\ & = &
\sum_{l}
\left<
{\cal U}_{m_1,l}(\tau)^\dag
{\cal U}_{m_2,n_1}(\tau) \wp_{l,n_2} \right>_\mathrm{stoch}
, \;\; \tau > 0 .
\nonumber
\end{eqnarray}
For $\tau < 0 $,
swap the upper and lower rows of subscripts on the left hand side,
as in Eq.~(\ref{Eq:wp2-stat-sym}).
In the double entropy basis this is
\begin{eqnarray} \label{Eq:trans-prob}
\wp^{(2),SS}_{\stackrel{\scriptstyle m_2,n_2}{m_1,n_1}}(\tau)
& = &
\left<
{\cal U}_{n_2,n_2}^S(\tau)^\dag
{\cal U}_{n_1,n_1}^S(\tau) \wp_{n_2,n_2}^S \right>_\mathrm{stoch}
\nonumber \\ &  & \mbox{ } \times
\delta_{m_2,n_1}  \delta_{m_1,n_2}
, \;\; \tau > 0 .
\end{eqnarray}
Recall that the probability operator and the time propagator
are diagonal in the entropy representation.

As a check that the time propagator expression
for the time correlation function is correct,
the leading order terms in the small $\tau$ expansion
of both sides of  Eq.~(\ref{Eq:trans-prob}) will now be obtained explicitly.
This is only done in the high temperature limit
in which the expectation entropy and the actual entropy are the same.
The right hand side is
\begin{eqnarray}
\mbox{RHS}^{SS}_{\stackrel{\scriptstyle m_2,n_2}{m_1,n_1}}
& = &
\frac{\delta_{m_2,n_1}  \delta_{m_1,n_2}}{Z}
\left< e^{S^S_{n_2,n_2}/k_\mathrm{B}}
{\cal U}^{S}_{n_2,n_2}(\tau)^\dag
\right. \nonumber \\ && \left. \mbox{ } \times
{\cal U}^{S}_{n_1,n_1}(\tau) \right>_\mathrm{stoch}
, \;\; \tau > 0.
\end{eqnarray}
Now with $\hat \lambda \equiv \sum_n \lambda_n
| \zeta_n^S \rangle \langle \zeta_n^S|$
and $\hat {\cal R} \equiv \sum_n r_n
| \zeta_n^S \rangle \langle \zeta_n^S|$,
the stochastic propagator is
\begin{eqnarray}
{\cal U}^{S}_{m,n}(\tau) & = &
\left\{ \hat{\mathrm I}
+ \frac{\tau}{i\hbar} \hat{\cal H}
- \frac{|\tau|}{2T\overline N} \hat\lambda
\left[ \hat{\cal H} - E_0 \hat{\mathrm I} \right]
+ \hat {\cal R}
\right\}_{m,n}^{S}
\nonumber \\ &=&
\left\{ 1
+ \frac{\tau}{i\hbar} E_n
- \frac{|\tau|}{2T\overline N} \lambda_n
\left[ E_n - E_0  \right]
\right. \nonumber \\ && \left. \mbox{ }
\rule{0cm}{.5cm} 
+ r_n
\right\} \delta_{m,n} .
\end{eqnarray}
Using this and the variance of the random operator,
to linear order in the time step
the right hand side of the transition probability,
Eq.~(\ref{Eq:trans-prob}), is
\begin{eqnarray} 
\lefteqn{
\mbox{RHS}^{SS}_{\stackrel{\scriptstyle m_2,n_2}{m_1,n_1}}
} \nonumber \\
& = &
\frac{\delta_{m_1,n_2}\delta_{m_2,n_1}}{Z}
e^{S^S_{n_1,n_1}/k_\mathrm{B}}
\nonumber \\ && \mbox{ } \times
\left\{
1 + \frac{\tau}{i\hbar} \left[ E_{n_1} - E_{n_2}\right]
\right. \nonumber \\ && \left. \mbox{ }
- \frac{|\tau|}{2T\overline N} \left(
\lambda_{n_1} \left[ E_{n_1} - E_0\right]
+ \lambda_{n_2} \left[ E_{n_2} - E_0\right] \right)
\right. \nonumber \\ && \left. \mbox{ }
+ \frac{|\tau| \lambda_{n_1} \left[ E_{n_1} - E_0\right]
}{2T\overline N} \delta_{m_1,n_1}
\right\} .
\end{eqnarray}
The double diagonal part of this is
\begin{equation} \label{Eq:trans-RHS-diag}
\mbox{RHS}^{SS}_{\stackrel{\scriptstyle m,m}{m,m}}
=
\frac{1}{Z}
e^{S^S_{m,m}/k_\mathrm{B}} .
\end{equation}
This is just the singlet probability.
These diagonal terms are independent of the choice of the $\lambda_n$.
For these the time correlation function
is independent of the specific model of the reservoir
interactions with the sub-system.

The non-diagonal terms are
the ones that depend upon the time step,
and this means that the time correlation function
as a function of $\tau$ will depend upon the drag coefficients.
However, since the reservoir should represent a weak perturbation
of the sub-system (the boundary region should be much smaller
than the sub-system itself),
the magnitudes of the $\lambda_n$ should be small enough
that the irreversible term (the one proportional to $|\tau|$)
should be dominated by the adiabatic, reversible term
(the one proportional to $\tau$).

In order to show that this is equal to the left hand side
of Eq.~(\ref{Eq:trans-prob}),
which are the coefficients of the transition probability matrix,
the eigenfunctions of the second entropy operator are required.
Recall that the domain is ${\mathrm H} \otimes {\mathrm H}$,
and so one can invoke a basis of the form
$\{ |\zeta_{m_2}\rangle ,  |\zeta_{m_1}\rangle \}$.
(One could normalize this by a factor of $\surd 2$.)
In quantum mechanics one can neglect an overall phase factor
for the wave function.
However in the present case there is the possibility
of a phase difference between the two Hilbert spaces
and this should be considered.
In particular the basis
$\{ -|\zeta_{m_2}\rangle ,  |\zeta_{m_1}\rangle \}$
differs from the first basis by a phase of $\pi$
and spans a space orthogonal  to that spanned by the first one.
In the following analysis wave functions will be projected onto the
sub-space spanned by the first basis on the grounds that it is
the dominant one.
The first reason for this is that the small time step limit is
being considered, in which case the phase of $\psi_2$
must be almost the same as that of $\psi_1$.
The second reason is that it will be shown
that the leading eigenfunction in the second basis
is  negative and diverges in the small time step limit.
Because this appears in the exponent
of the transition probability,
it contributes negligibly to the transition.

The second entropy operator acting
on a basis vector in the double entropy basis is
\begin{eqnarray}
\lefteqn{
\hat S^{(2)}(\tau)
\left( \begin{array}{c}
 |\zeta_{m_2}^S \rangle \\  |\zeta_{m_1}^S \rangle
\end{array} \right)
} \nonumber \\ & = &
2 \overline N
\left( \begin{array}{cc}
 \hat a(\tau) & \hat b(\tau)  \\  \hat b(\tau)^\dag & \hat c(\tau)
\end{array} \right)
\left( \begin{array}{c}
 |\zeta_{m_2}^S \rangle \\  |\zeta_{m_1}^S \rangle
\end{array} \right)
+
\overline S^{(1)}
\left( \begin{array}{c}
 |\zeta_{m_2}^S \rangle \\  |\zeta_{m_1}^S \rangle
\end{array} \right)
\nonumber \\ & = &
2 \overline N
\left( \begin{array}{cc}
\frac{-\hat \lambda^{-1} }{|\tau|}  + \hat a_0  + \widehat \tau a_0' &
\frac{\hat \lambda^{-1} }{|\tau|}  + \hat b_0    \\
\frac{\hat \lambda^{-1} }{|\tau|}  + \hat b_0  &
\frac{-\hat \lambda^{-1} }{|\tau|}  + \hat a_0  - \widehat \tau a_0'
\end{array} \right)
\left( \begin{array}{c}
 |\zeta_{m_2}^S \rangle \\  |\zeta_{m_1}^S \rangle
\end{array} \right)
\nonumber \\ &  & \mbox{ }
+ \overline S^{(1)}
\left( \begin{array}{c}
 |\zeta_{m_2}^S \rangle \\  |\zeta_{m_1}^S \rangle
\end{array} \right)
+ {\cal O}(\tau).
\end{eqnarray}
It is clear that to leading order
the eigenfunction are of the form
$| \zeta_m^S , \zeta_m^S \rangle$ with eigenvalue $\overline S^{(1)}$,
and $| -\zeta_m^S , \zeta_m^S \rangle$ with eigenvalue
$- 4 \overline N \lambda_m^{-1} /|\tau|$.
The latter is large and negative in the small time step limit,
and eigenfunctions of this form can be neglected.

One needs to go to the first correction to the leading eigenvalue
in order to obtain its dependence on the modes.
The next order eigenfunction has the form
\begin{equation}
\left( \begin{array}{c}
 |\zeta_{m}^S \rangle \\  |\zeta_{m}^S \rangle
\end{array} \right)
 +
\tau \left( \begin{array}{c}
-\hat \beta |\zeta_{m}^S \rangle \\ \hat \beta |\zeta_{m}^S \rangle
\end{array} \right)
+
|\tau| \left( \begin{array}{c}
-\hat \gamma |\zeta_{m}^S \rangle \\ \hat \gamma |\zeta_{m}^S \rangle
\end{array} \right) .
\end{equation}
With this the second order term, ${\cal O}(\tau^0)$,
for the right hand side of the eigenfunction equation is
\begin{eqnarray}
\mbox{RHS} & = &
4 \overline N \left( \begin{array}{c}
( \widehat \tau \hat \beta + \hat \gamma ) \hat \lambda^{-1}
|\zeta_{m}^S \rangle   \\
-( \widehat \tau \hat \beta + \hat \gamma ) \hat \lambda^{-1}
|\zeta_{m}^S \rangle
\end{array} \right)
+ \overline S^{(1)}
\left( \begin{array}{c}
 |\zeta_{m}^S \rangle \\  |\zeta_{m}^S \rangle
\end{array} \right)
\nonumber \\ && \mbox{ }
+
2 \overline N \left( \begin{array}{c}
( \hat a_0 + \hat b_0 + \widehat \tau \hat a_0'  )
|\zeta_{m}^S \rangle   \\
( \hat a_0 + \hat b_0 - \widehat \tau \hat a_0' )
|\zeta_{m}^S \rangle
\end{array} \right)
+ {\cal O}(\tau).
\end{eqnarray}
In order for this to be an eigenfunction,
$\hat \gamma $ must equal zero.
The coefficient $\widehat \tau \equiv \mbox{sign }\tau$ vanishes when
\begin{equation}
\hat \beta = \frac{-1}{2} \hat a_0' \hat \lambda .
\end{equation}
One now has, to the two leading orders,
\begin{eqnarray}
\mbox{RHS} & = &
\overline S^{(1)}
\left( \begin{array}{c}
 |\zeta_{m}^S \rangle \\  |\zeta_{m}^S \rangle
\end{array} \right)
+
2 \overline N \left( \begin{array}{c}
( \hat a_0 + \hat b_0   )
|\zeta_{m}^S \rangle   \\
( \hat a_0 + \hat b_0  )
|\zeta_{m}^S \rangle
\end{array} \right)
+ {\cal O}(\tau)
\nonumber \\ &=&
\overline S^{(1)}
\left( \begin{array}{c}
 |\zeta_{m}^S \rangle \\  |\zeta_{m}^S \rangle
\end{array} \right)
+ \overline N \left( \begin{array}{c}
\hat S'' |\zeta_{m}^S \rangle   \\
\hat S'' |\zeta_{m}^S \rangle
\end{array} \right)
+ {\cal O}(\tau)
\nonumber \\ &=&
\left[ \overline S^{(1)}
- \frac{E_m-E_0}{T}
\right]
\left( \begin{array}{c}
 |\zeta_{m}^S \rangle \\  |\zeta_{m}^S \rangle
\end{array} \right)
+ {\cal O}(\tau).
\end{eqnarray}
The the $m$-dependent part of the prefactor is the relevant eigenvalue,
$S_{mm}^S = - E_m/T$,
the constant remainder being incorporated into the
the normalizing partition function $Z^{(2)}(\tau)$.
Hence the double diagonal part of
the left hand side of Eq.~(\ref{Eq:trans-prob}) is
\begin{eqnarray}
\wp^{(2),SS}_{\stackrel{\scriptstyle m,m}{m,m}}(\tau)
& = &
\frac{1}{Z^{(2)}(\tau)}
\langle \zeta_{m}^S, \zeta_{m}^S |
e^{\hat S^{(2)} /k_\mathrm{B} }
| \zeta_{m}^S, \zeta_{m}^S \rangle
\nonumber \\ & = &
\frac{1}{Z'^{(2)}(\tau)} e^{S_{mm}^S /k_\mathrm{B} } .
\end{eqnarray}
One sees that this is equal to the right hand side,
Eq.~(\ref{Eq:trans-RHS-diag}),
which validates
the propagator expression for the
time correlation function, Eq.~(\ref{Eq:TCF-prop-1}).
It is not possible to carry out the check to ${\cal O}(\tau)$
because the expansion for the second entropy is only valid to
${\cal O}(\tau^0)$.

\subsubsection{Parity}

Operators that represent physical observables are Hermitian,
$\hat A = \hat A^\dag$.
Without loss of generality they may be taken to be either real or imaginary,
since, if complex, they can be split into their real and imaginary parts.
Let $\epsilon_A = \pm 1$ denote the parity of the operator,
$\hat A= \epsilon_A \hat A^* = \epsilon_A \hat A^\mathrm{T}$, and
similarly for other operators.

The expectation value 
in the wave state $\psi$ is
\begin{equation}
A(\psi) =
\frac{\underline{\underline A}: \underline \psi \, \underline \psi^*
}{\underline \psi^* \cdot \underline \psi} ,
\end{equation}
and that in  the wave state $\psi^*$ is
\begin{equation}
A(\psi^*)
= \frac{\underline{\underline A}: \underline \psi^* \underline \psi
}{\underline \psi \cdot \underline \psi^*}
= \frac{\underline{\underline A}^\mathrm{T} :
\underline \psi \, \underline \psi^*
}{\underline \psi^* \cdot \underline \psi}
= \epsilon_A A(\psi) .
\end{equation}
Since conjugation of the wave state represents velocity reversal,
one sees from this that
the parity of an operator
signifies whether it is even or odd under time reversal.

Since Hermitian operator have real expectation values,
$C_{BA}(0)$ is real if, and only if, $\hat B \hat A$ is Hermitian.
Since the two operators are individually Hermitian,
this implies that they must commute,
\begin{equation}
C_{BA}(0)^* = C_{BA}(0)
\Leftrightarrow
\hat B \hat A  = \hat A  \hat B.
\end{equation}
In this case $C_{BA}(0^+) = C_{BA}(0^-)$.

Explicitly one has
\begin{eqnarray}
C_{BA}(0^+)^*
& = &
\langle \hat B  \hat A \rangle_\mathrm{stat}^*
\nonumber \\ & = &
\int \mathrm{d} \psi \;
\frac{ \langle \psi| \hat \wp \hat B  \hat A |\psi \rangle^*
}{\langle \psi |\psi \rangle^* }
\nonumber \\ & = &
\int \mathrm{d} \underline \psi \;
\frac{1}{N(\psi)}
\underline \psi
\cdot \underline{\underline \wp}^*
\cdot \underline{\underline B}^*
\cdot \underline{\underline A}^*
\cdot \underline \psi^*
\nonumber \\ & = &
\epsilon_A  \epsilon_B
\int \mathrm{d} \underline \psi^* \;
\frac{1}{N(\psi)}
\underline \psi^*
\cdot \underline{\underline \wp}
\cdot \underline{\underline B}
\cdot \underline{\underline A}
\cdot \underline \psi
\nonumber \\ & = &
\epsilon_A  \epsilon_B C_{BA}(0^+) .
\end{eqnarray}
The final equality follows because the integral is over all
of Hilbert space, and so $\underline \psi^*$
is a dummy variable of integration.
Since the time correlation function is real if the operators commute,
this proves that
\begin{equation}
\left< \hat B(0) \hat A(0) \right>_\mathrm{stat} = 0 \mbox{ if }
\epsilon_A \ne \epsilon_B
\mbox{ and } \hat B \hat A  = \hat A  \hat B.
\end{equation}
In other words,
commuting operators with opposite parity are instantaneously uncorrelated.
If the operators do not commute and have opposite parity,
then the time correlation function at $\tau=0$
is imaginary, $C_{BA}(0^+)^* = - C_{BA}(0^+)$.
Hence one can say that in general
$\mbox{Re } C_{BA}(0) = 0 $ if $\epsilon_A \ne \epsilon_B$.

From the symmetry conditions for the second entropy given
in \S\ref{Sec:S2-flucn},
one sees that the complex conjugate of the transition probability operator
has the symmetry
\begin{equation}
\hat \wp^{(2)}(\tau)^* = \hat \wp^{(2)}(-\tau) .
\end{equation}
Accordingly
\begin{eqnarray}
C_{BA}(\tau)^* & = &
\mbox{Tr}^{(2)} \hat \wp^{(2)}(\tau)^* \{ \hat B^*,\hat A^* \}
\nonumber \\ & = &
\epsilon_A  \epsilon_B
\mbox{Tr}^{(2)} \hat \wp^{(2)}(-\tau) \{ \hat B,\hat A \}
\nonumber \\ & = &
\epsilon_A  \epsilon_B  C_{BA}(-\tau) .
\end{eqnarray}

If $\hat A$ and $\hat B(\tau)$ commute,
then their time correlation function is real,
$C_{BA}(\tau)^* = C_{BA}(\tau)$.
In this case
\begin{equation}
C_{BA}(\tau)
=
\epsilon_A  \epsilon_B C_{BA}(-\tau)
=
\epsilon_A  \epsilon_B C_{AB}(\tau) .
\end{equation}
This may also be seen
from the  propagator expression, Eq.~(\ref{Eq:TCF-prop-1}).
Noting that $\hat {\cal U}(\tau)^* = \hat {\cal U}(-\tau)$,
and assuming that the time correlation function  is real,
one has
\begin{eqnarray}
C_{BA}(\tau)  & = & C_{BA}(\tau)^*
\nonumber \\ & = &
\mbox{Tr}^{(1)} \left< \hat {\cal U}(\tau)^\dag
\hat B  \hat {\cal U}(\tau) \hat A \hat \wp \right>_\mathrm{stoch}^*
\nonumber \\ & = &
\mbox{Tr}^{(1)}
\left< \hat {\cal U}(\tau)^{\dag*}
\hat B^*  \hat {\cal U}(\tau)^* \hat A^* \hat \wp^* \right>_\mathrm{stoch}
\nonumber \\ & = &
\epsilon_A  \epsilon_B
\mbox{Tr}^{(1)}
\left< \hat {\cal U}(-\tau)^{\dag}
\hat B  \hat {\cal U}(-\tau) \hat A \hat \wp \right>_\mathrm{stoch}
\nonumber \\ & = &
\epsilon_A  \epsilon_B C_{BA}(-\tau) .
\end{eqnarray}
This result  is
the analogue
of the classical result given in \S 2.5.1 of Ref.~\onlinecite{NETDSM}.





\begin{thebibliography}{99}


\bibitem{TDSM}
Attard, P. (2002),
\emph{Thermodynamics and Statistical Mechanics:
Equilibrium by Entropy Maximisation},
(Academic Press, London).

\bibitem{NETDSM}
Attard, P. (2012),
\emph{Non-Equilibrium Thermodynamics and Statistical Mechanics:
Foundations and Applications},
(Oxford University Press, Oxford).

\bibitem{AttardIX}
Attard, P. (2009),
J. Chem.\ Phys.\ {\bf 130}, 194113.


\bibitem{Davies76}
Davies, E. B. (1976),
\emph{Quantum Theory of Open Systems},
(Academic Press, London). 

\bibitem{Kallianpur80}
Kallianpur, G. (1980),
\emph{Stochastic Filtering Theory},
(Springer, Berlin).

\bibitem{Gisin84}
Gisin, N. (1984),
Phys.\ Rev.\ Lett.\ {\bf 52},  1657. 

\bibitem{Belavkin89}
Belavkin, V. (1989),
Phys.\ Lett.\ A, {\bf 140},  355. 

\bibitem{Kummerer03}
K\"ummerer, B. and Maassen, H. (2003),
J. Phys.\ A {\bf 36},  2155. 

\bibitem{Bouten04}
Bouten, L., Gu\c t\v a, M., and Maassen, H. (2004),
J. Phys.\ A {\bf 37}, 3189. 

\bibitem{Pellegrini08}
Pellegrini, C. (2008)
Ann.\ Probab.\ {\bf  36},   2332. 



\bibitem{Breuer02}
Breuer, H.-P. and Petruccione, F. (2002),
\emph{The Theory of Open Quantum Systems},
(Oxford University Press, Oxford).  

\bibitem{Weiss12}
Weiss, U. (2012),
\emph{Quantum Dissipative Systems},
(World Scientific, New Jersey).


\bibitem{Dekker77}
Dekker, H. (1977),
Phys.\ Rev.\ A {bf 16}, 2116.

\bibitem{Kostin72}
Kostin, M. D. (1972),
J.\ Chem.\ Phys.\ {\bf 57}, 3589.

\bibitem{Yasue78}
Yasue, K. (1978),
Ann.\ Phys.\ N. Y. {\bf 114}, 479.

\bibitem{Nelson66}
Nelson, E. (1966),
Phys.\ Rev.\ {\bf 150}, 1079.



\bibitem{Percival98}
Percival, I. (1998),
\emph{Quantum State Diffusion},
(Cambridge University Press, Cambridge).

\bibitem{Wiseman92}
Wiseman, H. M. and Milburn, G. J. (1992),
Phys.\ Rev.\ A {\bf 47}, 1652.



\bibitem{Spohn80}
Spohn, H. (1980),
Rev.\ Mod.\ Phys.\ {\bf 52}, 569.

\bibitem{Gardiner91}
Gardiner, C. W. (1991),
\emph{Quantum Noise},
(Springer, Berlin).


\bibitem{Gardiner88}
Gardiner, C. W. (1988),
IBM J. Res.\ Dev.\ {\bf 32}, 127.

\bibitem{Iles-Smith13}
Iles-Smith,  J.,   Lambert, N., and   Nazir, A. (2013),
arXiv:1311.0016. 

\bibitem{Ford87}
Ford, G. W. and Kac, M. (1987),
J.\ Stat.\ Phys.\ {\bf 46},  803.

\bibitem{Ford88}
Ford, G. W., Lewis, J. T., and O'Connell, R. F. (1988),
Phys.\ Rev.\ A {\bf 37}, 4419.

\bibitem{Caldeira81}
Caldeira, A. O. and Leggett, A. J. (1981),
Phys.\ Rev.\ Lett.\ {\bf 46}, 211.


\bibitem{QSM1}
Attard, P.
arXiv:1401.1786  (2013).

\bibitem{Neumann27}
von Neumann, J. (1927),
G\"ottinger Nachrichten {\bf 1}, 245. 
von Neumann, J. (1932), \emph{Mathematische Grundlagen der
Quantenmechanik}, (Springer, Berlin).

\bibitem{Messiah61}
Messiah, A. (1961),
\emph{Quantum Mechanics},
(North-Holland, Amsterdam, Vols I and II).


\bibitem{Merzbacher70}
Merzbacher, E. (1970),
\emph{Quantum Mechanics},
(Wiley, New York, 2nd edn).

\bibitem{Bogulbov82}
Bogulbov, N. N. and Bogulbov, N. N. (1982),
\emph{Introduction to Quantum Statistical Mechanics},
(World Scientific, Singapore).


\bibitem{QSM3}
Attard, P. (2014),
arXiv:1404.2683.


\bibitem{Attard12}
Attard, P. (2012),
arXiv:1209.5500. 


\bibitem{AttardII}
Attard, P. (2005),
J. Chem.\ Phys.\ {\bf 122}, 154101.










\end{thebibliography}
\end{document}